\documentclass[preprint2,trackchanges]{aastex6}
\usepackage{xspace}
\usepackage{graphicx}
\newcommand*{\suzaku}{{\it Suzaku}\xspace}
\newcommand*{\os}{OAO\,1657-415\xspace}
\newcommand*{\go}{GX\,301-2\xspace}
\newcommand*{\gx}{GX\,1+4\xspace}
\newcommand*{\vl}{Vela\,X-1\xspace}

\newcommand*{\alfven}{Alfv\'en\xspace}
\newcommand*{\ka}{\mbox{${\rm K_{\alpha}}$}\xspace}
\newcommand*{\kb}{\mbox{${\rm K_{\beta}}$}\xspace}

\begin{document}


\title{The Modulating Optical Depth of Photoelectric Absorption Edge with Pulse Phase in Accretion-Powered X-ray Pulsars}

%% Use \author, \affil, plus the \and command to format author and affiliation 
%% information.  If done correctly the peer review system will be able to
%% automatically put the author and affiliation information from the manuscript
%% and save the corresponding author the trouble of entering it by hand.
%%
%% The \affil should be used to document primary affiliations and the
%% \altaffil should be used for secondary affiliations, titles, or email.

%% Authors with the same affiliation can be grouped in a single
\author{Yuki Yoshida\altaffilmark{1,2}}
\and
\author{Shunji Kitamoto\altaffilmark{1,2}}

%% Use the \and command so offset the last author.
%\and
\altaffiltext{1}{Department of Physics, College of Science, Rikkyo University, 3-34-1 Nishi-Ikebukuro, Toshima, Tokyo 171-8501, Japan}
\altaffiltext{2}{Research Center for Measurement in Advanced Science, Rikkyo University, 3-34-1 Nishi-Ikebukuro, Toshima, Tokyo 171-8501, Japan}
\email{yy@rikkyo.ac.jp}

\begin{abstract}
We report the first discovery of pulse phase modulating optical depths at the iron K-edge in accretion-powered X-ray pulsars, from \suzaku observations. 
A significant modulating optical depth of the iron K-edge is detected for \vl and \gx. 
Similar trends are seen in \go and \os, though with poor statistical significance.
The observed iron K-edge exhibits a maximum optical depth, when the X-ray continuum dims,
and there is no significant pulse phase variation in the ionization state of iron.
The revealed changes in the optical depth with pulse phase can be explained as being due to the accreting matter captured by the magnetic field lines of the pulsar, which co-rotates with the neutron star spin and is responsible for photoelectric absorption. 
Based on the above interpretation, we propose that the accreting matter within the \alfven radius contains iron with an ionization state of Fe$_{\rm V\hspace{-0.1em}I{\mbox \textendash}X\hspace{-0.1em}I}$ with a particle density of $10^{17}$~cm$^{-3}$ and has a shape flattened along the azimuthal direction, such as an accretion curtain.
\end{abstract}




%%%%%%%%%%%%%%%%%%%%%%%%%%%%%%%%%%%%%%%%%%%%%%%%%%%%%%%%%%%%%%
\section{Introduction\label{sec:introduction}}
%%%%%%%%%%%%%%%%%%%%%%%%%%%%%%%%%%%%%%%%%%%%%%%%%%%%%%%%%%%%%%
The X-ray emission lines and photoelectric absorption edges, especially of iron, that are often observed in the spectra of accretion-powered X-ray pulsars \citep[e.g.,][]{White:1983aa}, 
can be attributed to the X-ray source and its vicinity and are reflective of the physical properties of the plasma.
Therefore, measuring these spectral features can provide information on the physical state of the X-ray source, its vicinity and surrounding matter.
The iron emission line at 6.4~keV from accretion-powered X-ray pulsars has been considered to be X-ray fluorescence emitted from neutral or low-ionized iron illuminated by continuous X-rays from the neutron star or its vicinity \citep[e.g.,][]{Koyama:1985aa,Makishima:1986aa}.
The absorption edge of iron has also been detected in association with the emission line, 
and the origin of this has been considered to be the same as the fluorescence lines based on the ionization state estimated from the energies of the edge \citep{Makishima:1986aa,Ebisawa:1996aa,Watanabe:2006aa}.
It is generally difficult to determine an emission source or an emission region of emission lines, 
though we can restrict the location of matter responsible for the absorption edge to a region between the X-ray source and the observer. There are other interpretations, however, such as reflective processes.
Thus, if the origin of the absorption edge is clarified, the matter distributed around the neutron star can be identified more definitely than a determination based only on the information obtained from the emission lines.
A detailed investigation of the absorption edge feature is therefore an interesting tool for studying accretion flow around neutron stars.

The time variation of these spectral features gives various information, such as the size or geometry of the reprocessing region.
In accretion-powered X-ray pulsars, the emission line flux sometimes exhibits variability depending on the pulse phase \citep[e.g.,][]{Leahy:1990aa,Day:1993aa,Choi:1996aa,Vasco:2013aa}. 
However, sometimes no modulation with the pulse period is found \citep[e.g.,][]{Ohashi:1984aa,Paul:2002aa,Lei:2009aa,Suchy:2012aa}. 
On the other hand, modulation of the optical depth of the absorption edge with the pulse period in X-ray pulsars has never been reported. Pulse phase variations of the absorption column density, which are derived from the low energy cutoff, have been seen in a few accretion powered X-ray pulsars (KS\,1947+300;\citealt{Ballhausen:2016aa}; \gx; \citealt{Galloway:2001aa}).

Up until the end of its observational operation in 2015,  \suzaku observed more than 30 accretion-powered X-ray pulsars,  60 times or more in total.
All of the acquired data are available in an archive.
From this data we selected pulsars with a spin period longer than 30~s for a phase-resolved analysis.
We then selected statistically good datasets, excluding data sets with short exposure time less than 10~ks and faint sources with less than 0.5~count\,s$^{-1}$ on each X-ray imaging spectrometer \citep[XISs;][]{Koyama:2007aa},
giving about 30 available data sets. We extracted the time-averaged spectra from the data.
Finally we selected four data sets of four accretion-powered X-ray pulsars, \go, \vl, \os, and \gx, because their spectra distinctly exhibit the K-shell emission line and absorption K-edge of iron around 6.4~keV and 7.1~keV even without spectral model fitting, as shown in Figure~\ref{fig:eg_broadband_spec}.

The \suzaku observations of the four pulsars provide the opportunity to investigate the time variability in both the iron line emissions and the iron absorption edge through phase-resolved spectroscopy thanks to its high sensitivity and excellent energy resolution, as well as their long spin periods.
In this paper, we report the results of the detailed spectral analysis of the \suzaku observations. We present the first discovery of modulation in the optical depth of the iron K-edge with the pulsar spin period.
These results reveal the physical conditions of the matter surrounding the neutron stars in the systems.

%%%%%%%%%%%%%%%%%%%%%%%%%%%%%%%%%%%%%%%%%%%%%%%%%%%%%%%%%%%%%%
\section{Observation and Data Reduction\label{sec:observation}}
%%%%%%%%%%%%%%%%%%%%%%%%%%%%%%%%%%%%%%%%%%%%%%%%%%%%%%%%%%%%%%
%%============================================================================================
\subsection{Observation\label{subsec:observation}}
%%============================================================================================
Table~\ref{tab:obs_lst} summarizes the details of the \suzaku observations of the four selected X-ray pulsars.
In all the observations, the XISs were operated in the normal mode incorporating a 1/4 window option, which gives a time resolution of 2~s during the observation. 
In the observation of \os, spaced-row charge injection (SCI) was performed with 2~keV equivalent electrons for XIS\,0 and XIS\,3 (front-illuminated or FI CCDs) and 6~keV equivalent electrons for XIS\,1 (back-illuminated or BI CCD),
while in the other observations SCI was conducted with 2~keV equivalent electrons for both FI and BI CCDs. 
The hard X-ray detector \citep[HXD;][]{Takahashi:2007aa} was operated in the standard mode wherein individual events were recorded with a time resolution of 61~$\mu$s in all observations.

The total effective exposure times were more than 60~ks in all four observations.
In the case of \go, another \suzaku observation was conducted in 2008 with a total effective exposure time of 10~ks (ObsID=403044010).
We did not use this data because of its poor statistics for phase-resolved analysis with a spin period of 700~s.

%%============================================================================================
\subsection{Data Reduction\label{subsec:reduct}}
%%============================================================================================
The archival \suzaku data of the four sources were analyzed using HEASARC software version 6.25.
Only XIS\,0, 1, and 3 were used and analyzed, since XIS\,2 was out of operation due to damage by a micrometeorite in 2006 November.
The HXD-PIN and HXD-GSO data were used for broad-band spectral analysis. 
The HXD-PIN data were also used to determine the spin period of the pulsars, since they have better time resolution than XIS.
We reprocessed the XIS and HXD data using the FTOOLS task {\tt aepipeline} (version 2.3.12.25) with the calibration database (version hxd-20110913, xis-20181010 and xrt-20110630) and the standard screening criteria.

The on-source and background XIS events were extracted from a circular region with 3\arcmin radius and an annulus with an inner radius of 4\arcmin and an outer radius 7\arcmin centered at the source position, respectively. 
For \os and \gx, a pileup estimation following \citet{Yamada:2012aa} showed that the maximum pileup fraction at the center of the source region is less than 1\%, 
and the effect of the pileup was therefore neglected for these observations. 
On the other hand, from the on-source events of \go and \vl we excluded the events in central regions with radii of 0.\arcmin3 and 0.\arcmin9, respectively, to eliminate the pileup effect.
This procedure reduces the pileup fraction to below 1\%.
The redistribution matrix files and ancillary response files for each XIS detector were generated using the {\tt xisrmfgen} and {\tt xissimarfgen} routines \citep{Ishisaki:2007aa}, respectively. 
We used tuned (LCFITDT) non X-ray background (NXB) models \citep{Fukazawa:2009aa} to obtain the NXB events for HXD. 
From the HXD-PIN data, we subtracted the NXB and an expected contribution from the cosmic X-ray background (CXB) with a spectral shape given by \citet{Boldt:1987aa}. 
From the HXD-GSO data, we subtracted only the NXB, as the expected contribution of the CXB is negligible. 
In our analysis, response files released between 2008 January and 2011 June were used for the HXD-PIN data, selecting a suitable file for each observation, 
while response and effective area files released in 2010 May were used for the HXD-GSO data.

In the following analyses, all uncertainties quoted are given at the 90\% confidence level for the parameter of interest unless stated otherwise.

%%%%%%%%%%%%%%%%%%%%%%%%%%%%%%%%%%%%%%%%%%%%%%%%%%%%%%%%%%%%%%
\section{Analysis and Results\label{sec:analysis}}
%%%%%%%%%%%%%%%%%%%%%%%%%%%%%%%%%%%%%%%%%%%%%%%%%%%%%%%%%%%%%%
%%============================================================================================
\subsection{Light Curves and Pulse Period\label{subsec:timing}}
%%============================================================================================
For the timing analysis, we applied a barycentric correction to the arrival times of individual photons using the {\tt aebarycen} task of FTOOLS \citep{Terada:2008aa}. 
As for HXD events, dead-time corrections were applied using an FTOOLS task {\tt hxddtcor}. 
After the barycentric and dead-time corrections, we conducted a further correction for the effect of the orbital motion in the binary system 
with the orbital parameters reported by \citet{Koh:1997aa}, \citet{Kreykenbohm:2008aa}, and \citet{Jenke:2012aa} for \go, \vl, and \os, respectively. 
Light curves were extracted from the corrected XIS event data in the range 0.5--12~keV with a 2~s time-bin, which is the minimum available time resolution for the XIS operation mode (i.e., 1/4 window option). 
Data from all the operated XISs were added together and a single light curve was obtained. 
We also created light curves in the range 12--70~keV from HXD-PIN event data with a resolution of 1~s. 
From the above light curves, we subtracted both the NXB and the CXB.

\begin{figure}[ht]
\centering
\includegraphics[bb= 0 0 801 437,width=0.5\textwidth]{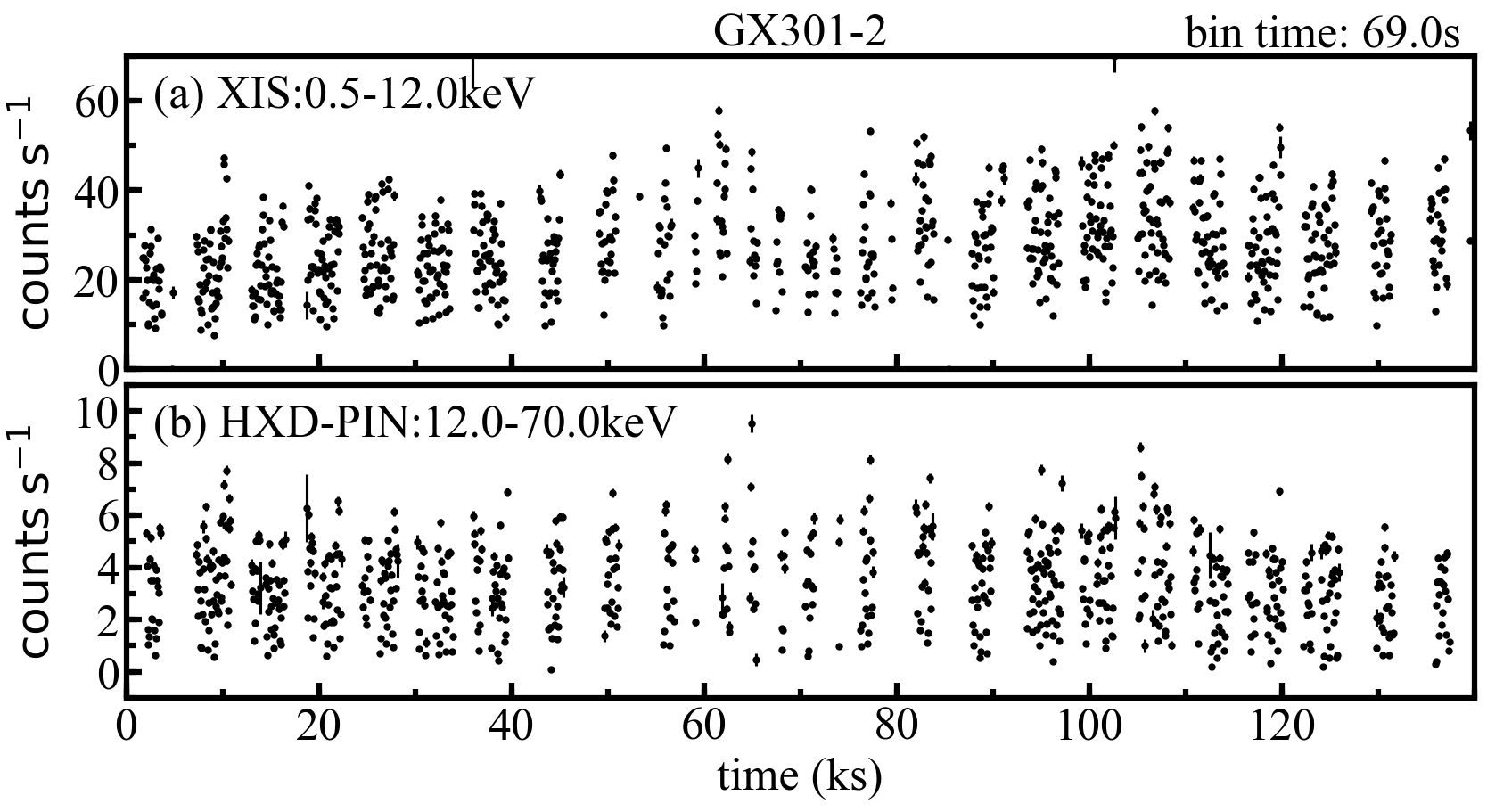} \\
\vspace{1em}
\includegraphics[bb= 0 0 801 437,width=0.5\textwidth]{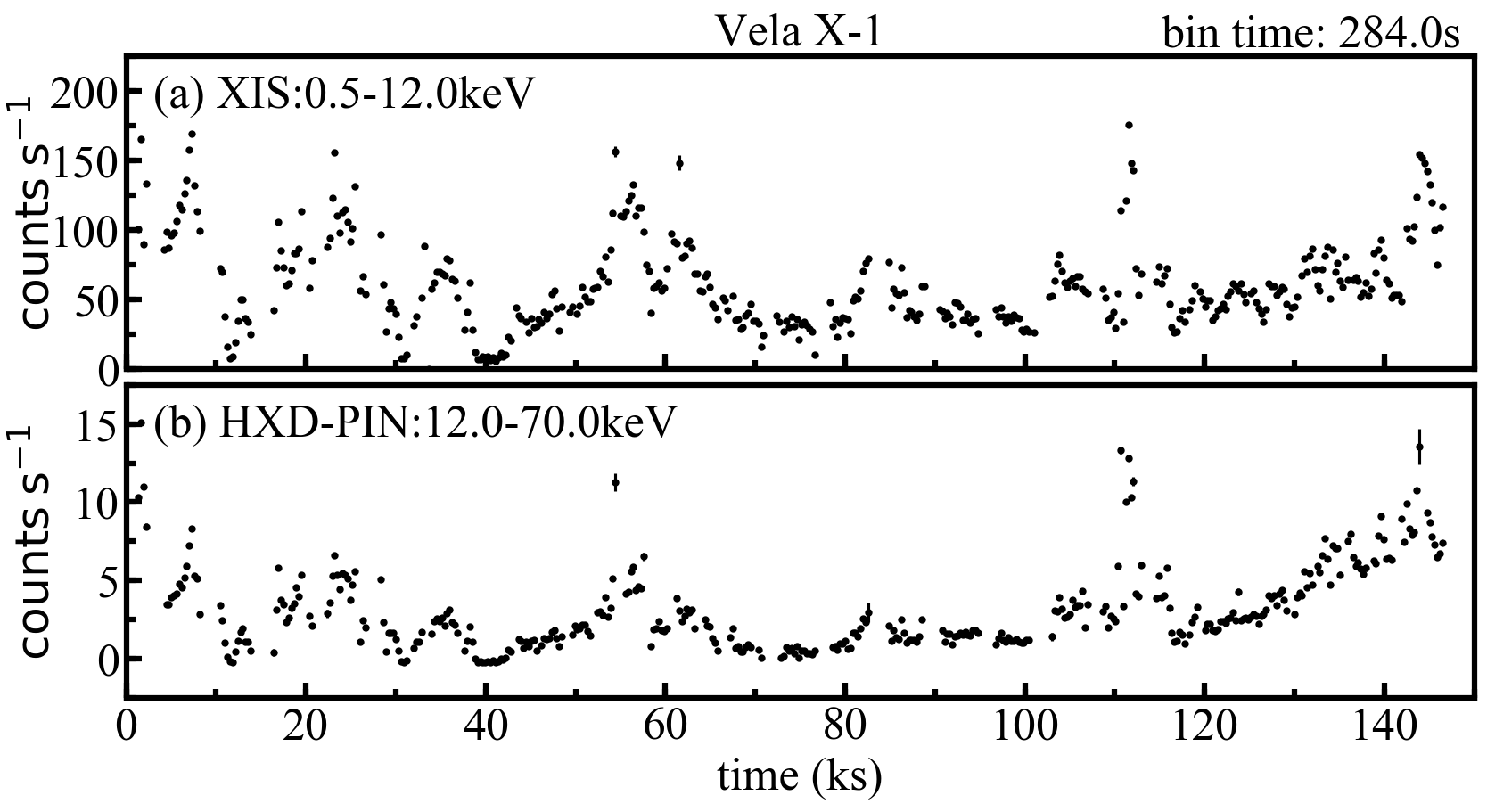} \\
\vspace{1em}
\includegraphics[bb= 0 0 801 437,width=0.5\textwidth]{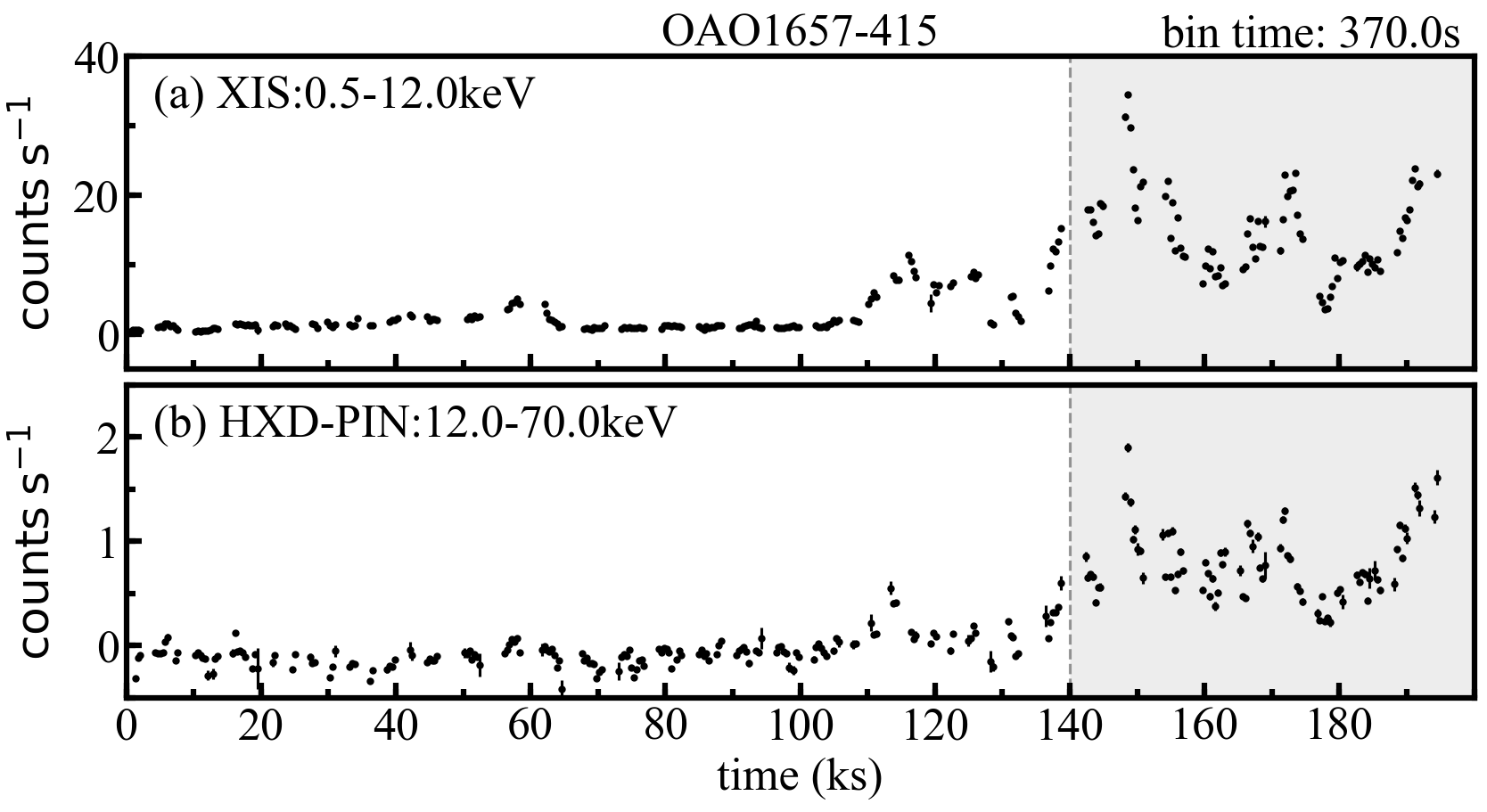} \\
\vspace{1em}
\includegraphics[bb= 0 0 801 437,width=0.5\textwidth]{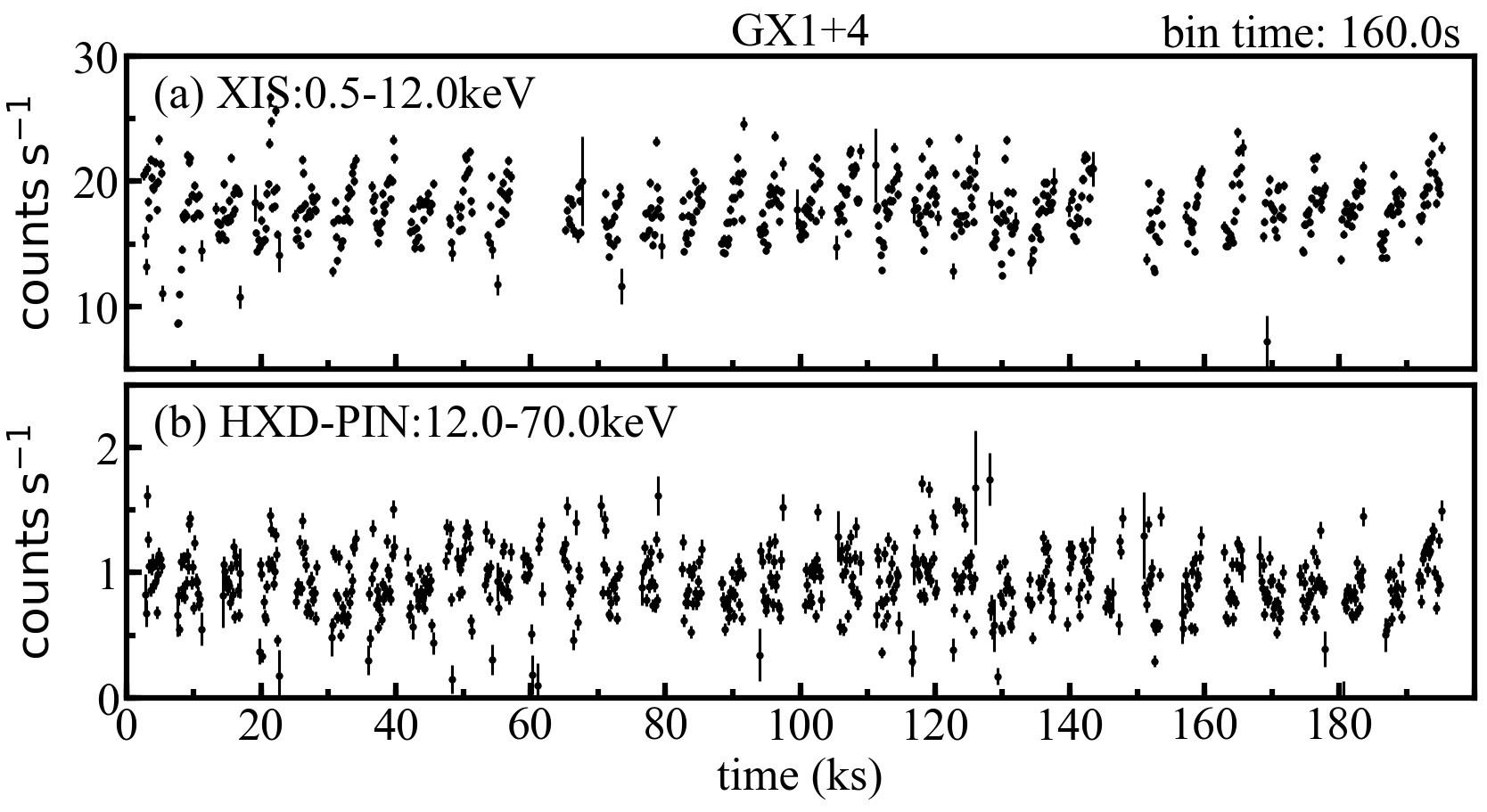}
\caption{
Background-subtracted, (a) XIS (0.5--12~keV; data added from three sensors) and (b) HXD-PIN (12--70~keV) light curves of \go, \vl, \os, and \gx, from top to bottom, binned with the times indicated above the panels. 
In the panel showing the light curves of \os, the gray-shaded regions indicate the duration of the high intensity period selected.
\label{fig:eg_xis_pin_lc}}
\end{figure}

Figure~\ref{fig:eg_xis_pin_lc} shows background-subtracted light curves of the four sources obtained from XIS (upper panel) and HXD-PIN (lower panel),
where the time-bins are tuned to the pulsar spin period or its harmonic value.
A prominent change in the intensity of the light curve of \os was found during the observation. 
A low intensity period was observed during the first $\sim1.4\times10^{5}$~s and a high intensity period followed for $\sim6\times10^{4}$~s. 
Since a change of pulse fraction with intensity was reported by \citet{Pradhan:2014aa} for the same \suzaku observation,
we divided the data into low and high intensity periods at $1.4\times10^{5}$~s from the start of the observation  
and performed separate analyses for the two periods.
However, the data in the low intensity period were not sufficiently statistically sound for a phase-resolved analysis and we could not extract any meaningful results.   
In the following description, we show only the results obtained from the last $\sim5\times10^{4}$ seconds of the observation, when the source exhibited high intensity. 
The selected time interval is shown in Figure~\ref{fig:eg_xis_pin_lc} by the shaded area.

To search for pulsations in the light curves of the sources for the phase-resolved analysis, 
we performed a standard epoch folding analysis \citep{Leahy:1983aa} of the HXD-PIN data for each observation with the {\tt efsearch} task of FTOOLS.
The barycentric pulsation periods observed in our analysis are summarized in Table~\ref{tab:TimAveResult}.
Errors estimated according to \citet{Leahy:1987aa} are also given in the table.
All the sources are consistent with the previous reported values  derived from the same \suzaku data \citep{Suchy:2012aa,Maitra:2013aa,Pradhan:2014aa,Yoshida:2017aa},  within the errors.

Using the estimated pulse periods, the light curves of the XISs were folded by applying the {\tt efold} task of FTOOLS.
Energy-divided pulse profiles, obtained from background subtracted light curves in the 0.5--7.1~keV and 7.1--12.0~keV ranges 
are shown in panels (a) and (b) of Figure~\ref{fig:eg_go_paramWphase}--\ref{fig:eg_gx_paramWphase},
where the data from the three XISs are summed.
According to the criteria described in \S~\ref{subsubsec:spectroscopy}, 
for phase-resolved spectroscopy, the data are divided into pulse phase intervals as indicated by the quoted numbers and overlaying colors in the same panels.

%%============================================================================================
\subsection{Time-averaged Spectrum\label{subsec:average}}
\begin{figure*}[ht]
\centering
\includegraphics[bb=0 0 576 432,width=0.5\textwidth]{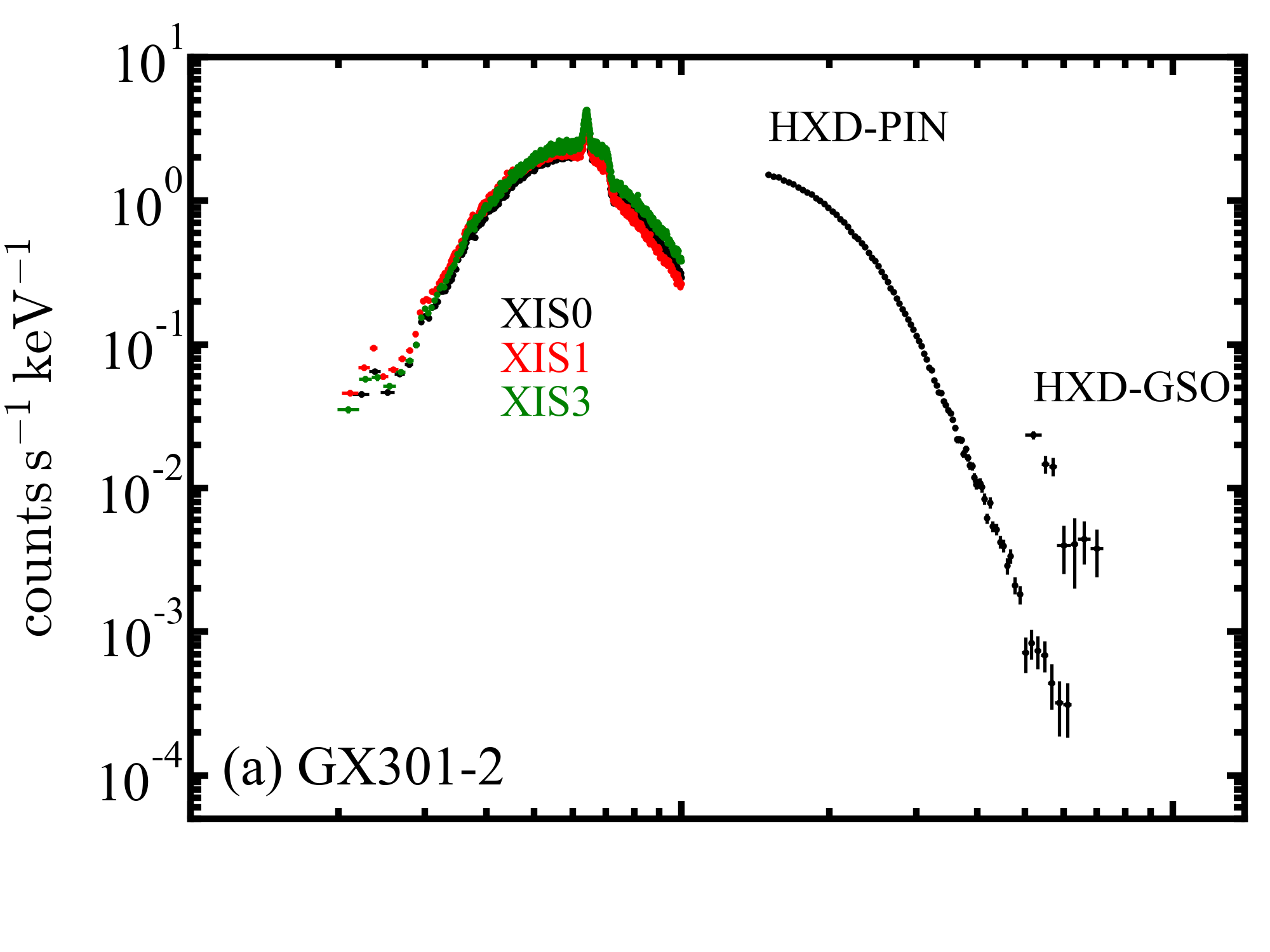}
\hspace{-5em}
\includegraphics[bb=0 0 576 432,width=0.5\textwidth]{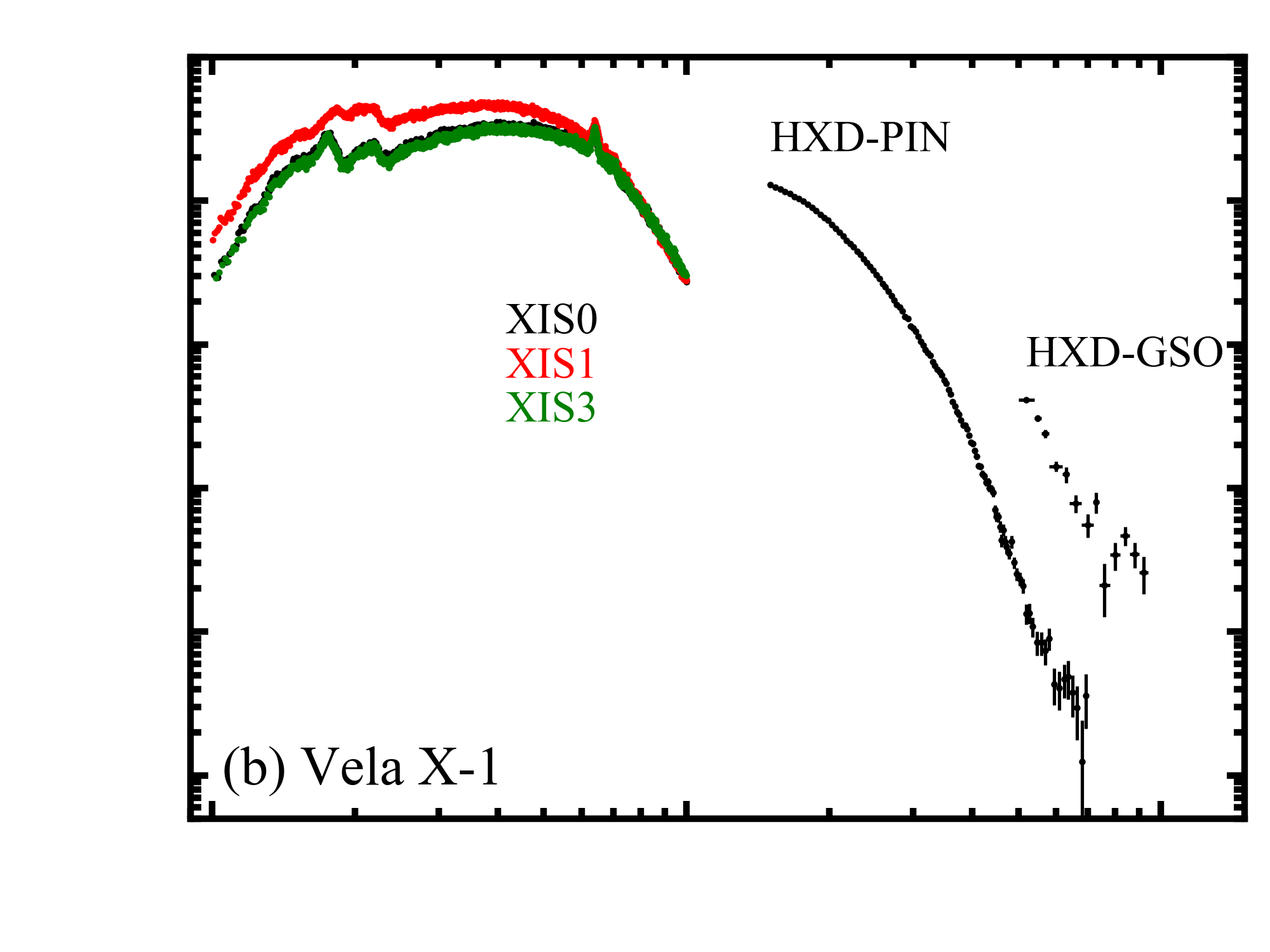} \\
\vspace{-4em}
\includegraphics[bb=0 0 576 432,width=0.5\textwidth]{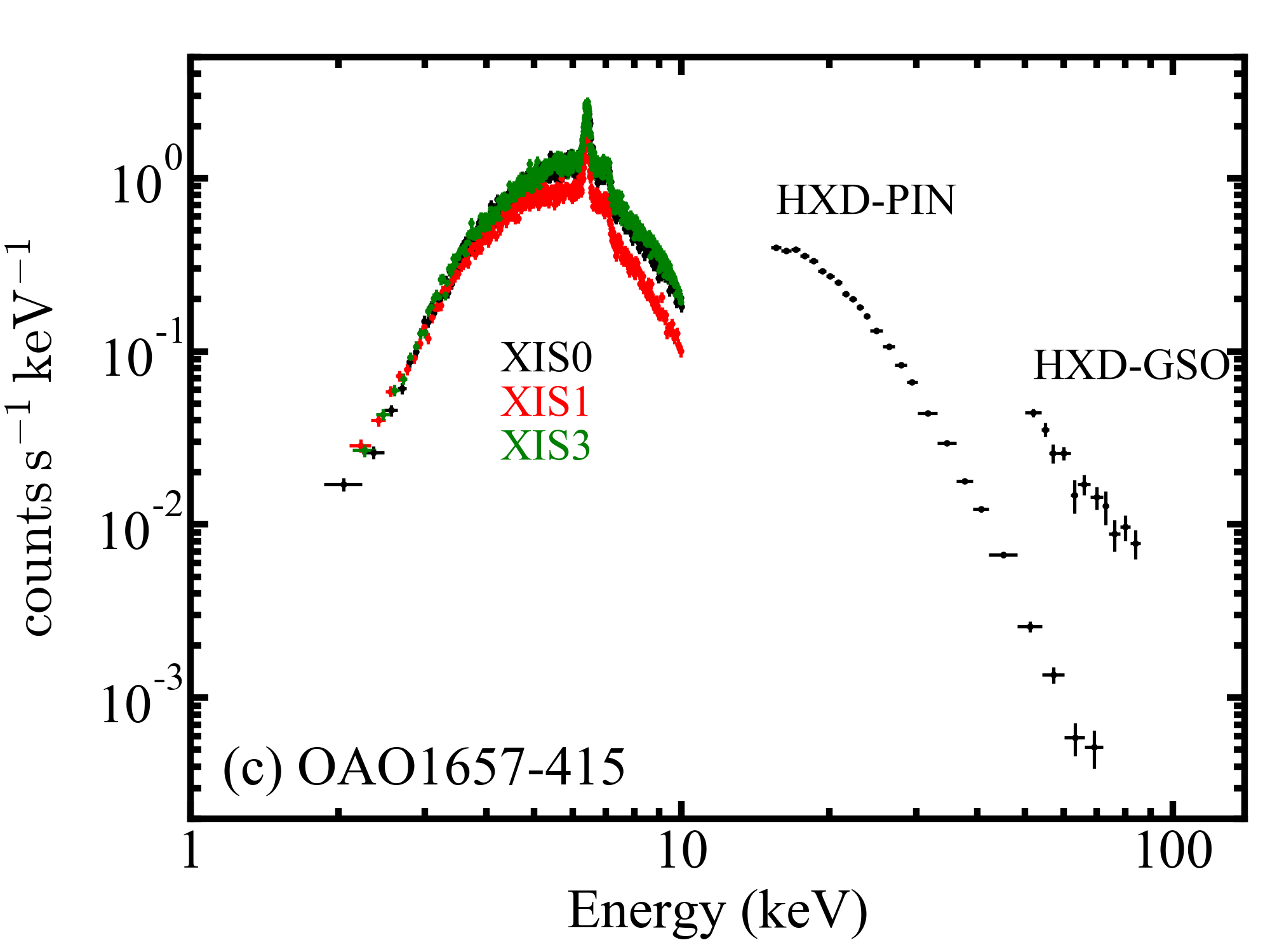}
\hspace{-5em}
\includegraphics[bb=0 0 576 432,width=0.5\textwidth]{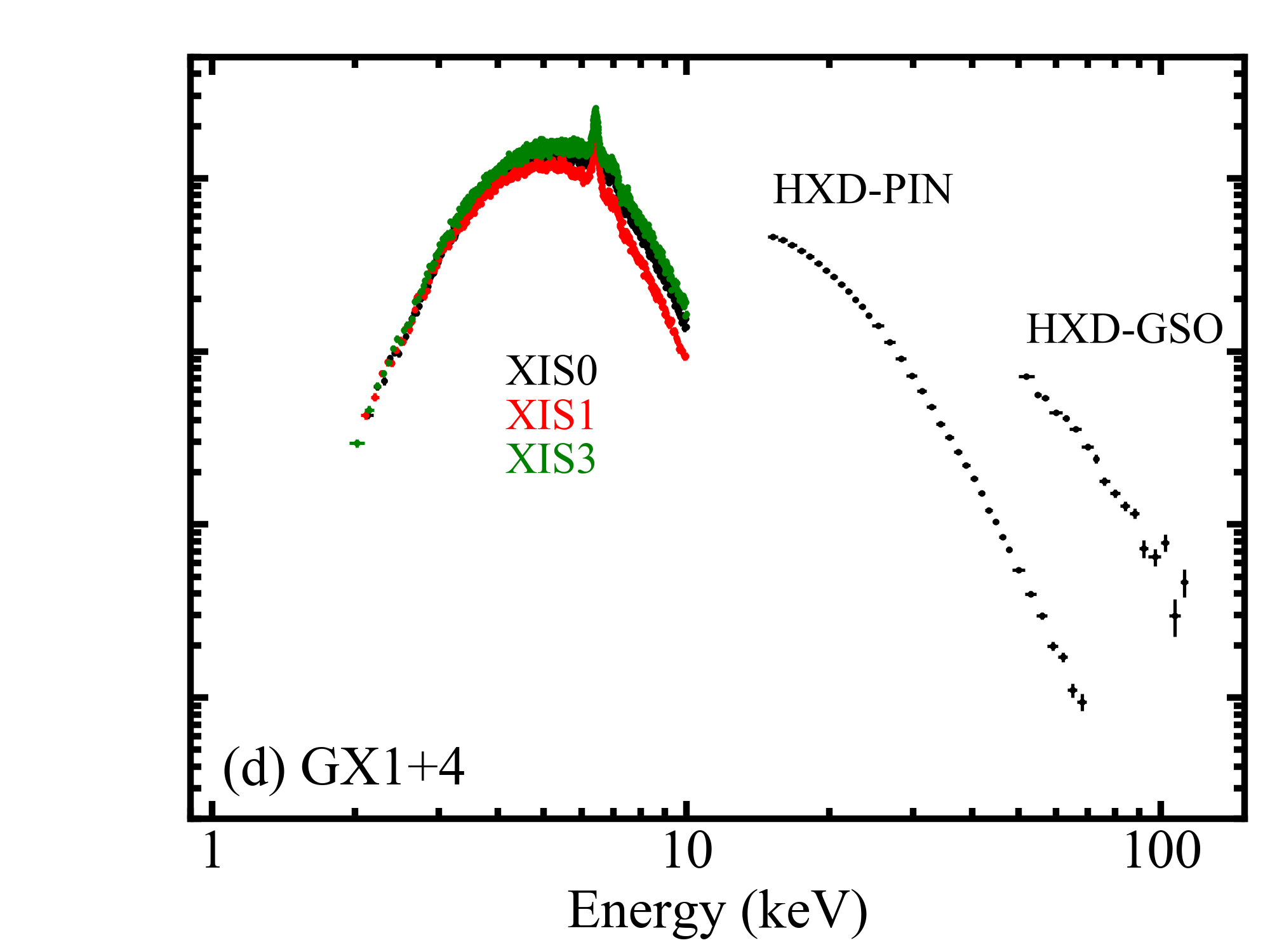}
\caption{Phase-averaged, background-subtracted spectra of \go (a), \vl (b), \os (c), and \gx (d), plotted in count rate space, 
obtained with XIS\,0 (black), XIS\,1 (red),  XIS\,3 (green), HXD-PIN, and HXD-GSO.
\label{fig:eg_broadband_spec}}
\end{figure*}

%%============================================================================================
Figure~\ref{fig:eg_broadband_spec} shows time-averaged and background-subtracted spectra of the four sources, though the spectrum of \os was obtained from the data of the high intensity period as described in \S~\ref{subsec:timing}. 
The data of \vl obtained by the three XISs was analyzed in the range 1--10~keV, 
while for the other cases, we used only 2--10~keV data for FI CCDs.  
Because they show strong low energy absorption (more than $N_{\rm H}=10^{23}$\,H\,atoms~cm$^{-2}$), as seen in Figure~\ref{fig:eg_broadband_spec}, 
the detected events below 2~keV are dominated by the ``low energy tail'' component \citep{Matsumoto:2006aa}, 
which is characteristic of the instruments and is not well calibrated \citep{Suchy:2012aa}. 
Data of XIS in the 2.2--2.4~keV energy range was ignored for the spectral fitting because of the presence of a gold edge feature, the calibration precision of which is not sufficient compared to the statistical uncertainty.  
We also omitted the 10--15~keV energy band of the HXD-PIN data because of the calibration uncertainty of the temperature dependent electrical noise.  
The cross normalizations of the XIS detectors were set to be free to cope with the calibration uncertainty of the effective area, 
whereas the cross normalizations between XIS and HXD were fixed at 1.16 and 1.18 for ``XIS nominal'' and ``HXD nominal'' observations, respectively, following the recommendation in the \suzaku ABC guide.

Using the XSPEC version 12.10 package \citep{Arnaud:1996aa}, these XIS and HXD spectra were simultaneously fitted by the model employed by earlier studies, to describe the same \suzaku broad-band spectra \citep{Suchy:2012aa,Pradhan:2014aa,Maitra:2013aa,Yoshida:2017aa}.
The spectral continuous components used are summarized in Table~\ref{tab:TimAveResult}.
A power-law multiplied Fermi--Dirac cutoff \citep[FDPL;][]{Tanaka:1986aa} model was applied to \go, the Negative and Positive power-law with an EXponential cutoff \citep[NPEX][]{Mihara:1995aa} model was used for \vl and \os, 
and an exponential cutoff power-law with a blackbody (BB$+$CPL) model was used for \gx.
All the spectral models were multiplied by the photoelectric absorption {\tt TBnew} \citep{Wilms:2000aa} with an improved abundance {\tt wilm} and cross section tables {\tt vern} \citep{Verner:1996aa}.
In addition, the model was modified by applying a partial covering intrinsic absorption \citep[e.g.,][]{Endo:2000aa}, except for the case of \gx.
To represent the cyclotron resonance scattering features in the spectra of \go and \vl, the multiplicative components {\tt gabs} and {\tt cyclabs} were introduced to the model.
In the spectra of \vl and \gx, we noticed absorption-edge-like residuals of data from the model around 7~keV. 
We fitted this edge-like feature by allowing the iron abundance of the absorber relative to the interstellar medium, $Z_{\rm Fe}$, to vary, and values of $Z_{\rm Fe}$ $\sim12$ in \vl and $\sim1.4$ in \gx were typically obtained. 
The obtained large abundance of \vl will be discussed in \ref{subsubsec:spectroscopy}.

The results of the time-averaged spectral fitting are summarized in Table~\ref{tab:TimAveResult}.
Using the obtained best-fit parameters, we estimated the X-ray luminosities of the four pulsars during the observations (for \os, only during the high intensity period), listed in Table~\ref{tab:TimAveResult}.
The distances to the sources from the observer are also given in the table.
These values are consistent with those reported by \citet{Suchy:2012aa}, \citet{Doroshenko:2011aa}, \citet{Jaisawal:2014aa}, and \citet{Yoshida:2017aa} for the same \suzaku data.

%%============================================================================================
\subsection{Phase-resolved Analysis\label{subsec:resolve}}
%%============================================================================================
%%============================================================================================
\subsubsection{Spectroscopy\label{subsubsec:spectroscopy}}
\begin{figure}[ht]
\centering
\includegraphics[bb= 0 0 601 906,width=0.5\textwidth]{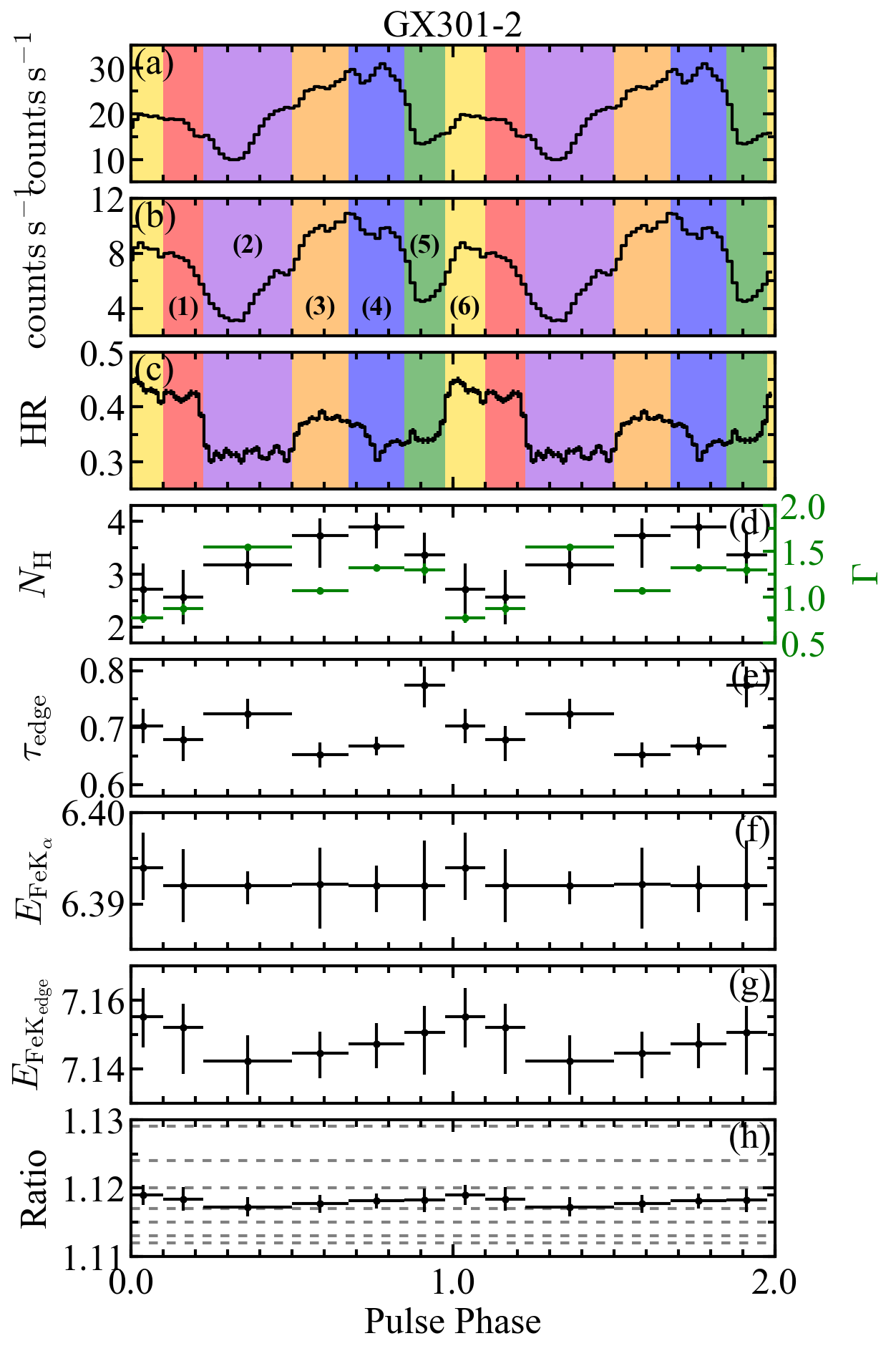}
\caption{Folded light curves and parameters of phase-resolved spectroscopy of \go with data during whole observation. 
Energy resolved, background-subtracted epoch folded light curves in the ranges (a) 0.5--7.1~keV and (b) 7.1--12.0~keV.
(c) Ratio between the above epoch folded light curves in two energy bands.
For phase-resolved spectroscopy, the data are divided into six pulse phase intervals, as indicated by the numbers and overlaying colors.
The spectral parameters are also plotted as a function of pulse phase. 
(d) Equivalent hydrogen column density (10$^{23}$\,H\,atoms~cm$^{-2}$) obtained from the low energy cutoff of the fully covering absorption component, plotted in black, and photon index of the power-law component, plotted in green, obtained by broad-band spectroscopy. 
 (e) Optical depth of iron K-edge, (f) center energies of the iron \ka line (in keV), and  (g) energies of the iron K-edge (in keV), derived from the fitting in the 5.0--7.9 keV energy band.
(h) Energy ratios of the iron \ka line to the iron K-edge.
The vertical axis on the right of panel (d) indicates the photon index of the power-law component. 
The gray dashed lines in panel (h) indicate the energy ratios for Fe$_{\rm I}$--Fe$_{\rm V\hspace{-0.1em}I\hspace{-0.1em}I}$ (from top to bottom).
Two cycles are shown for clarity. 
The errors of the spectral parameters shown here are for the 90\% confidence levels. 
 \label{fig:eg_go_paramWphase}}
\end{figure}

\begin{figure}[ht]
\centering
\includegraphics[bb= 0 0 601 906,width=0.5\textwidth]{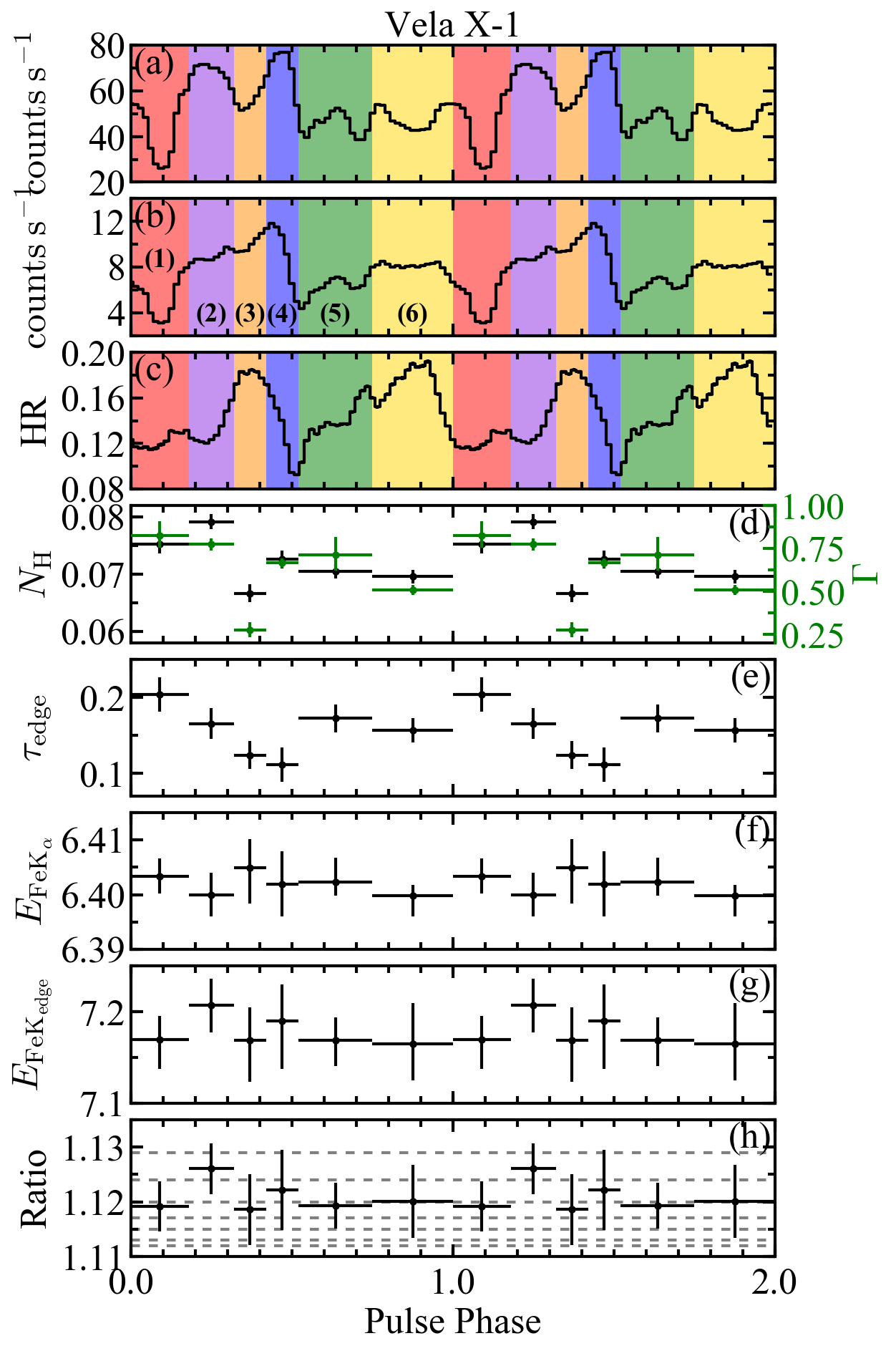}
\caption{As for Figure~\ref{fig:eg_go_paramWphase}, but with the results for \vl with data during the whole observation.
\label{fig:eg_vl_paramWphase}}
\end{figure}

\begin{figure}[ht]
\centering
\includegraphics[bb= 0 0 601 906,width=0.5\textwidth]{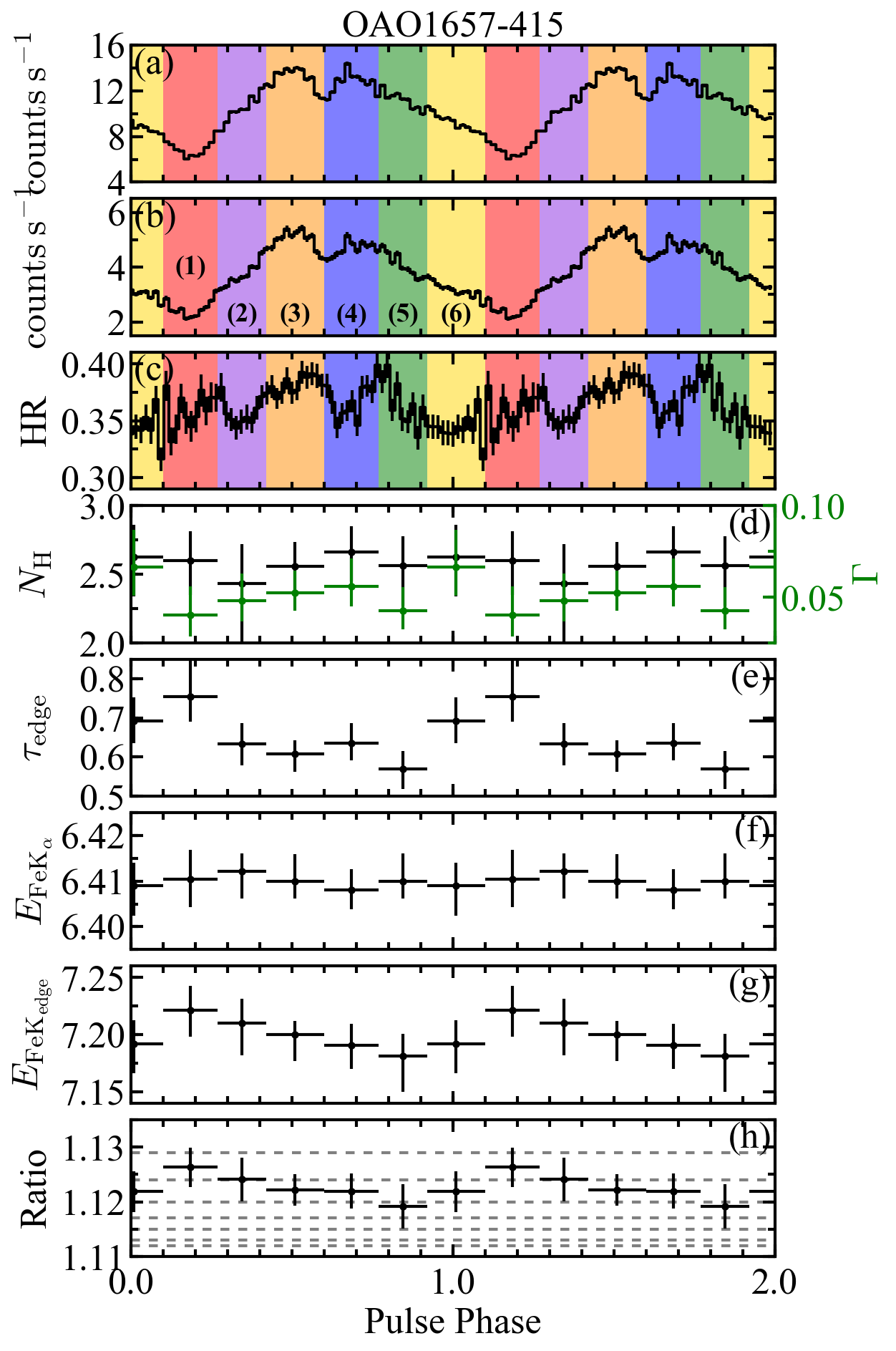}
\caption{As for Figure~\ref{fig:eg_go_paramWphase}, but with the results for \os with data during the high intensity period.
\label{fig:eg_os_timDiv_paramWphase}}
\end{figure}

\begin{figure}[ht]
\centering
\includegraphics[bb= 0 0 601 906,width=0.5\textwidth]{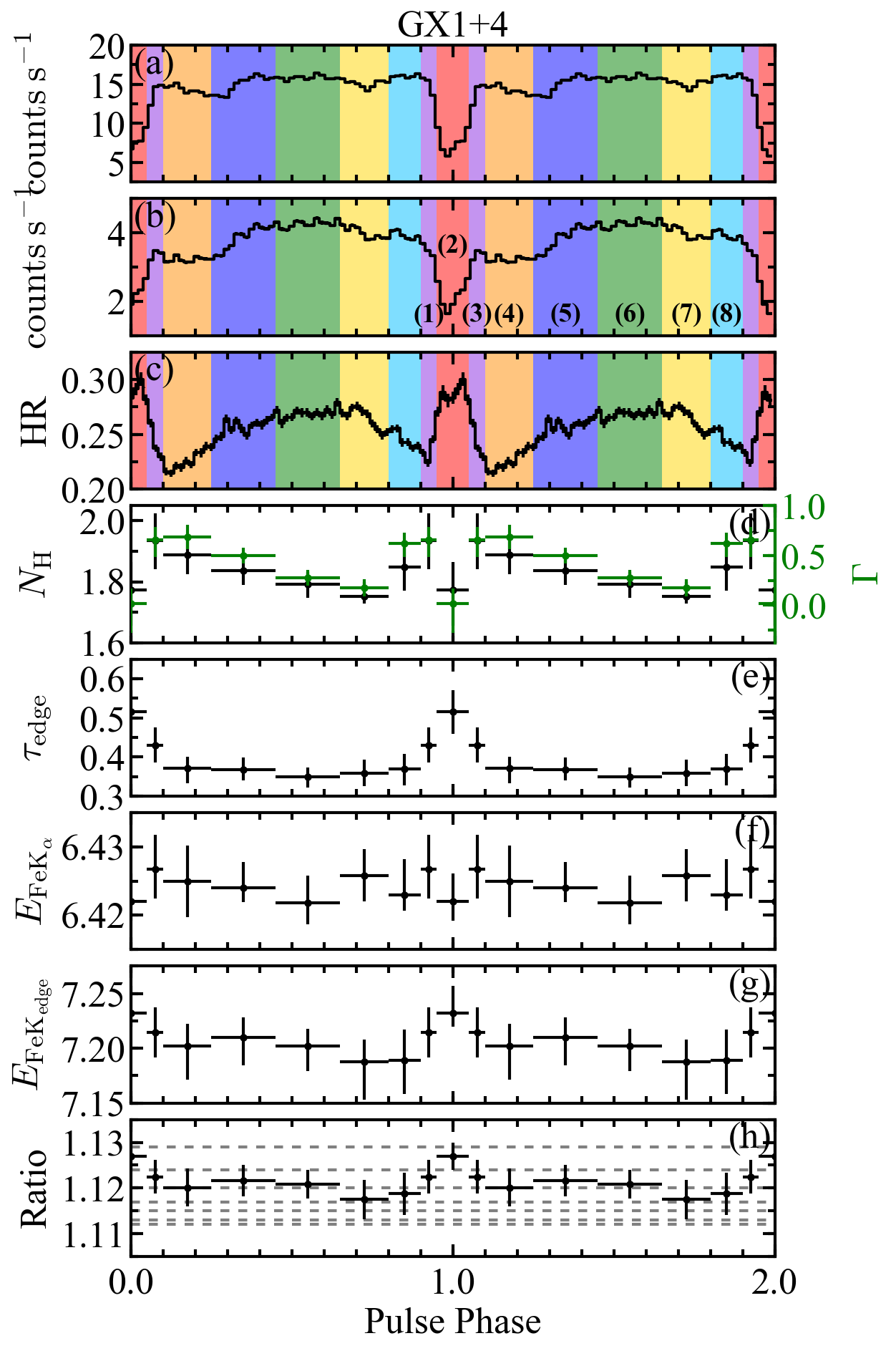}
\caption{As for Figure~\ref{fig:eg_go_paramWphase}, but with the results for \gx with data during the whole observation.
\label{fig:eg_gx_paramWphase}}
\end{figure}

%%============================================================================================
We performed phase-resolved spectroscopic analyses of the four sources, focusing on the emission lines and absorption edge of iron. 
To study the variation of the spectral shape according to the pulse phase, we divided the observation data into several phase bins, as indicated in Figures~\ref{fig:eg_go_paramWphase}--\ref{fig:eg_gx_paramWphase} with different overlaid colors. 
The bins were chosen to pick up various prominent features, e.g., the deep minimum and maximum peaks, 
as well as to retain photon statistics sufficient to constrain the individual spectral parameters. 
The hardness ratios between the two pulse profiles plotted in panel~(c) of Figures~\ref{fig:eg_go_paramWphase}--\ref{fig:eg_gx_paramWphase} were examined and the boundaries were found to be correct for picking up a stable hardness ratio region, except for continuously changing phases.
In the case of \gx, the phase intervals 1 and 3 (marked in purple in the figure) were combined to improve the photon statistics. 

We fitted the broad-band phase-resolved spectra with the same models as those used in the time-averaged spectral fitting listed in Table~\ref{tab:TimAveResult}. 
When the phase-resolved spectra of \go were fitted using the FDPL model, we fixed the cutoff energies at 29~keV, which is the best-fit value of the time-averaged spectrum, 
whereas the other parameters were left free. 
In the fitting of \vl and \gx, $Z_{\rm Fe}$ was fixed at the value obtained from the time-averaged fitting.
We found that the adopted model fitted all the phase-resolved spectra well. 

The best-fit parameters of the column density of the photoelectric absorption (of the fully covering component for \go, \vl, and \os) and the photon index (of the CPL model, negative power-law component in the NPEX model, and the FDPL model), derived from the above fitting are plotted as a function of the pulse phase in panel (d) of Figures~\ref{fig:eg_go_paramWphase}--\ref{fig:eg_gx_paramWphase}. 
The figures show strong positive-correlations between the column density and the photon index
suggesting that the derived values influence each other parameter. 
For example, if the spectral shape is steep in the high energy band, the derived column density value becomes high because the extrapolation to the low energy band of the steep spectrum requires a large photoelectric absorption, without any evidence to justify the extrapolation. 
Therefore, the absorption column densities are not physically meaningful values, since we cannot apply any physically well justified model in the low energy region.
If the column density is incorrect, the \ka line flux and the K-edge optical depth of the iron ions are also affected and we may derive wrong values.
We consider that the iron abundance of 12$\pm$2 of \vl obtained from the time-averaged spectral fit, shown in Table~\ref{tab:TimAveResult}, is wrong.

For a robust analysis of the variation of the emission lines and absorption edge with the pulse phase, avoiding the influence from the insufficient modeling of the broad-band spectrum, 
we restricted the energy range to 5.0--7.9~keV for the spectral fitting. 
In this restricted region, the continuum can be fitted by a power-law model multiplied by an absorption edge and hence we can deduce the edge parameter independently from the low-energy absorption.
An approximate model, {\tt edge}, is employed to reproduce the photoelectric absorption edge, which is given as follows:
\begin{eqnarray}
M(E)=
\left\{
   \begin{array}{ll} 
        1  & (E<E_{\rm c}) \\
        \exp\left(-\tau_{\rm edge}\left(\frac{E}{E_{\rm c}}\right)^{-3}\right) & (E\geqslant E_{\rm c}), 
   \end{array}
\right.\label{eq:edge} 
\end{eqnarray}
where $E_{\rm c}$ is the threshold energy and $\tau_{\rm edge}$ is the absorption optical depth at the threshold energy. 
The applied model consists of ${\tt powerlaw}\times{\tt edge}+{\tt gaussian}+{\tt gaussian}+{\tt gaussian}$, 
in which three {\tt gaussian} models express the iron \ka, iron \kb, and nickel \ka emission lines.
Since the iron \kb line contaminates the absorption edge feature, 
it is difficult to independently determine the parameters corresponding to these structures. 
Since the energy of the observed iron \ka emission line is around 6.4~keV, the ionization state of the iron should be low (Fe$_{\rm X\hspace{-0.1em}V\hspace{-0.1em}I\hspace{-0.1em}I}$ at most).
In the above ionization state, the difference of the ratio between the energies of the iron \ka and iron \kb lines is at most 1\%, 
but that of the energy of the iron edge to that of iron \ka line is at most 7\% \citep{Kallman:2004aa,Yamaguchi:2014aa}. 
Therefore, we fixed the ratio of the energy of the iron \kb line to that of the iron \ka line at 1.103 for the neutral case, 
while the energy of the iron K-edge was allowed to vary in the fitting.
In the case of \os, as the center energy of the nickel \ka line was poorly constrained due to low effective exposure in each phase bin, 
the ratio of its energy was fixed to that of the iron \ka line at 1.168 for the neutral case. 
The widths of the emission lines could not be resolved in the XIS data, and they were fixed to be sufficiently narrow.
Note that for our broad-band spectroscopy, we used only the FI CCDs for \go, \os, and \gx, 
because of a discrepancy between the FI and BI CCDs in the low energy region (below 3~keV). 
However, in the above restricted narrow energy range, the discrepancy between the FI and BI CCDs is negligibly small,
and we therefore used data from all the CCDs. 

Parameters obtained from the spectral fitting are summarized in Table~\ref{tab:phasedivfit} and plotted as a function of pulse phase in Figures~\ref{fig:eg_go_paramWphase}--\ref{fig:eg_gx_paramWphase}. 
We find that the optical depths of the iron K-edge, $\tau_{\rm edge}$, exhibit modulations with pulse phase, as seen in panel (e) of each figure. 
For all the four pulsars, $\tau_{\rm edge}$ increases at the phase when the X-ray continuum flux becomes dim. 
The energies of the iron \ka line and iron K-edge remain constant within the errors, as shown in panels (f) and (g). 
The ratios of the energy of the iron K-edge to that of the iron \ka line are calculated and plotted against the pulse phase in panel (h), 
and the values are constant within the errors. 
Note that for \os and \gx, the energy ratio shows marginal modulation due to the change of the energy of the edge,
though with poor statistical significance, which peaks synchronously with the maximum $\tau_{\rm edge}$. 
The expected energy ratios for Fe$_{\rm I}$--Fe$_{\rm V\hspace{-0.1em}I\hspace{-0.1em}I}$ are also shown in the same panels by gray dashed lines.
The averaged values of the energy ratios of the four pulsars are in the range 1.117--1.122. 
These values show that the ionization state of the iron is Fe$_{\rm I\hspace{-0.1em}I\mbox{\textendash}V\hspace{-0.1em}I\hspace{-0.1em}I}$. 
We confirmed that the assumed ratio of the energy of the iron \kb line to that of the iron \ka line of 1.103 in the fitting is consistent with this ionization state. 
Table~\ref{tab:phasedivfit} lists the obtained intensities of the iron \ka and \kb. 
The ratio between these intensities are roughly consistent with the expected values in the  above ionization state of 0.12--0.14 \citep{Yamaguchi:2014aa},  
except for the results of \go and of interval 3 of \os.
In these cases, the assumed energy ratio between the iron \ka and \kb lines might be smaller than the actual value, 
and thus the results for the edge parameters of \go and \os might be marginal. 
However, in the case of \vl and \gx, the edge parameters are not affected by the contamination of the iron \kb line. 

Using the obtained optical depth, an equivalent hydrogen column density for an assumed solar abundance of $N_{\rm H}=\frac{\tau_{\rm edge}}{Z_{\rm Fe}\sigma_{\rm Fe}}$ is derived, where $Z_{\rm Fe}$ and $\sigma_{\rm Fe}$ are the iron abundance in the solar system \citep[$=3.2\times10^{-5}$;][]{Asplund:2009aa} and the photoelectric absorption cross section of Fe$_{\rm I\hspace{-0.1em}I}$ at the energy of the K-edge \citep[$=3.3\times10^{-20}$~${\rm cm}^{2}$;][]{Verner:1995aa}.
The estimated equivalent hydrogen column densities for each phase bin are given in Table~\ref{tab:phasedivfit}.

%%============================================================================================
\subsubsection{Investigation of Statistical Significance of Modulation\label{subsubsec:stat}}
\begin{figure}[ht]
\centering
\includegraphics[bb= 0 0 593 713,width=0.5\textwidth]{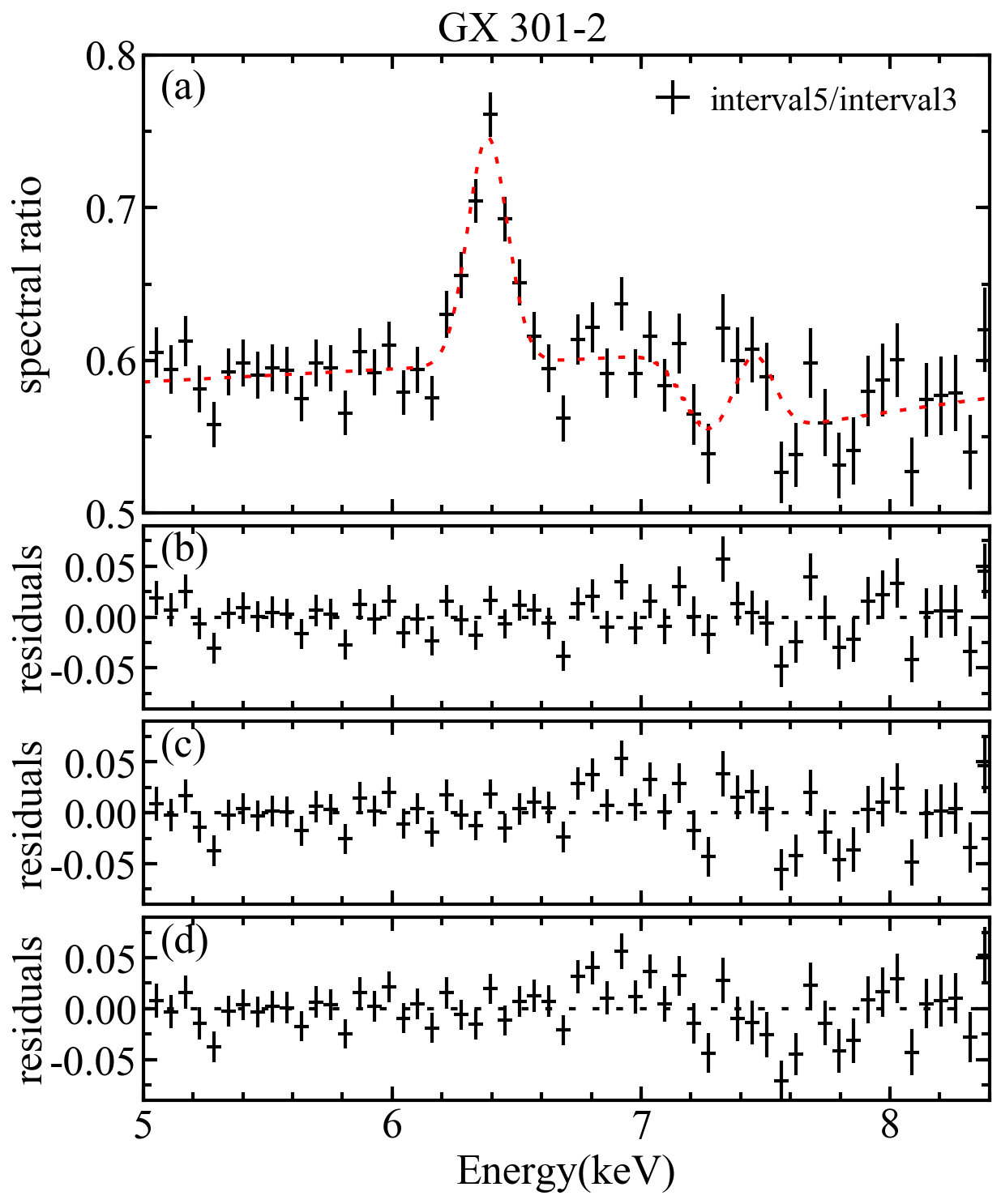}
\caption{
Spectral ratio between two phase-resolved spectra with the fitting results of \go with data during the whole observation.
(a) Spectral ratio between two phase-resolved spectra with the best-fit model of Equation~\ref{eq:edgefit}. 
(b) Residuals from the best-fit model in panel (a). 
(c), (d) As for panel (b), but the model excludes the edge component. The normalization of $G_{\rm Ni}(E)$ is kept free in panel (c) and is fixed at the value obtained from the fit using a model with an edge component in panel (d).
 \label{fig:eg_go_sprfit}}
\end{figure}

\begin{figure}[ht]
\centering
\includegraphics[bb= 0 0 593 713,width=0.5\textwidth]{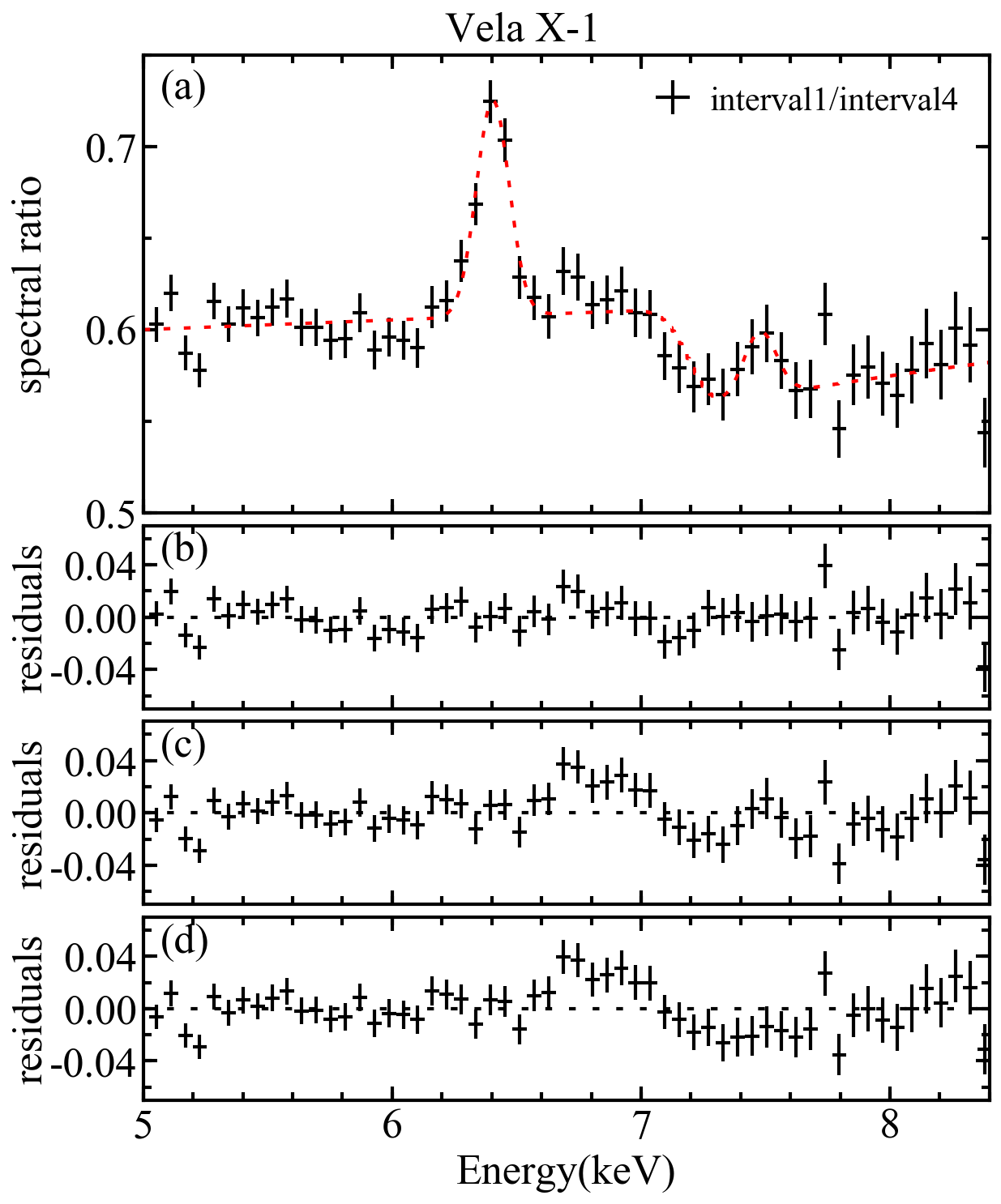}
\caption{As for Figure~\ref{fig:eg_go_sprfit}, but with the results for \vl with data during the whole observation.
\label{fig:eg_vl_sprfit}}
\end{figure}

\begin{figure}[ht]
\centering
\includegraphics[bb= 0 0 593 713,width=0.5\textwidth]{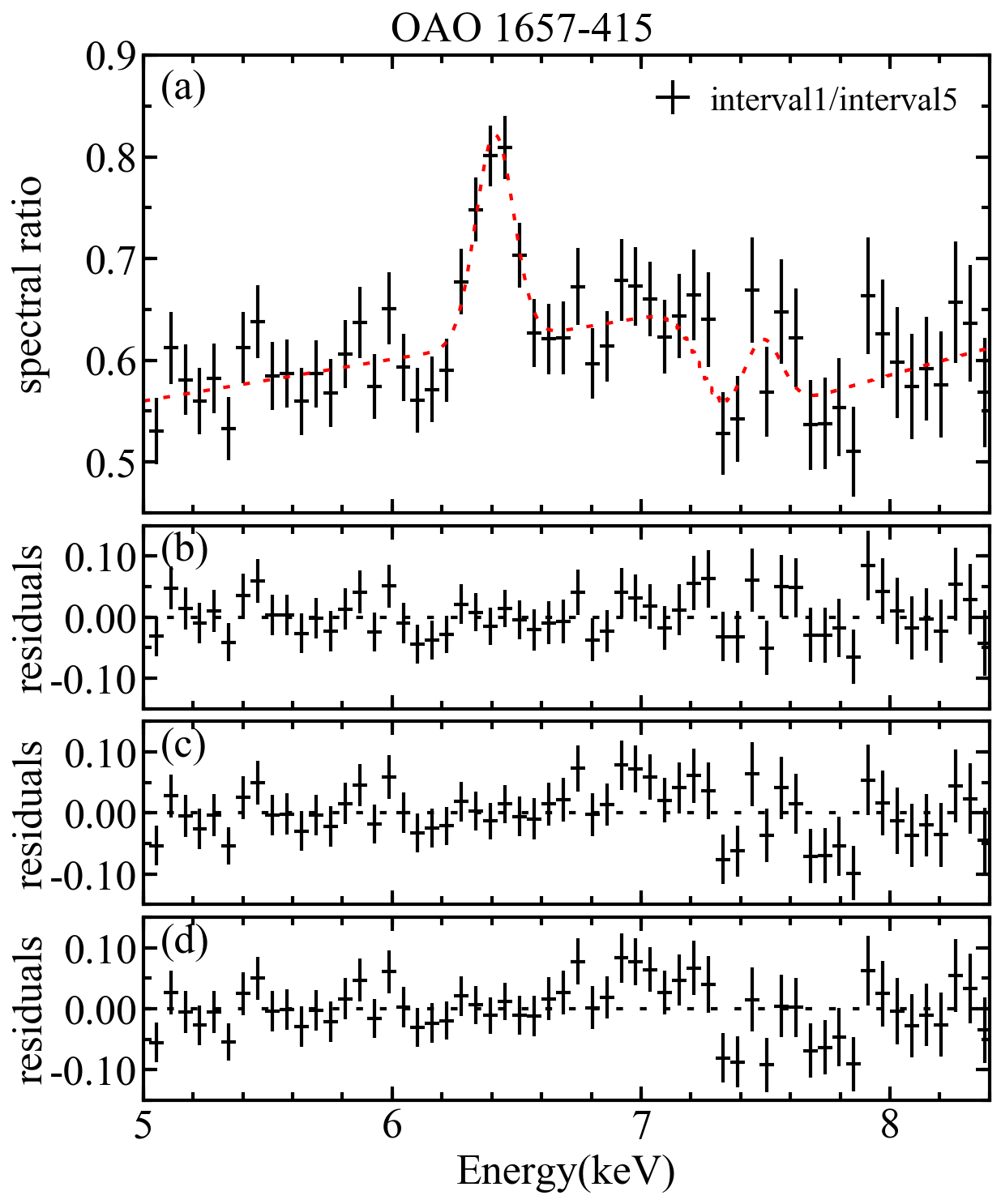}
\caption{As for Figure~\ref{fig:eg_go_sprfit}, but with the results for \os with data during the high intensity period.
\label{fig:eg_os_timDiv_sprfit}}
\end{figure}

\begin{figure}[ht]
\centering
\includegraphics[bb= 0 0 593 713,width=0.5\textwidth]{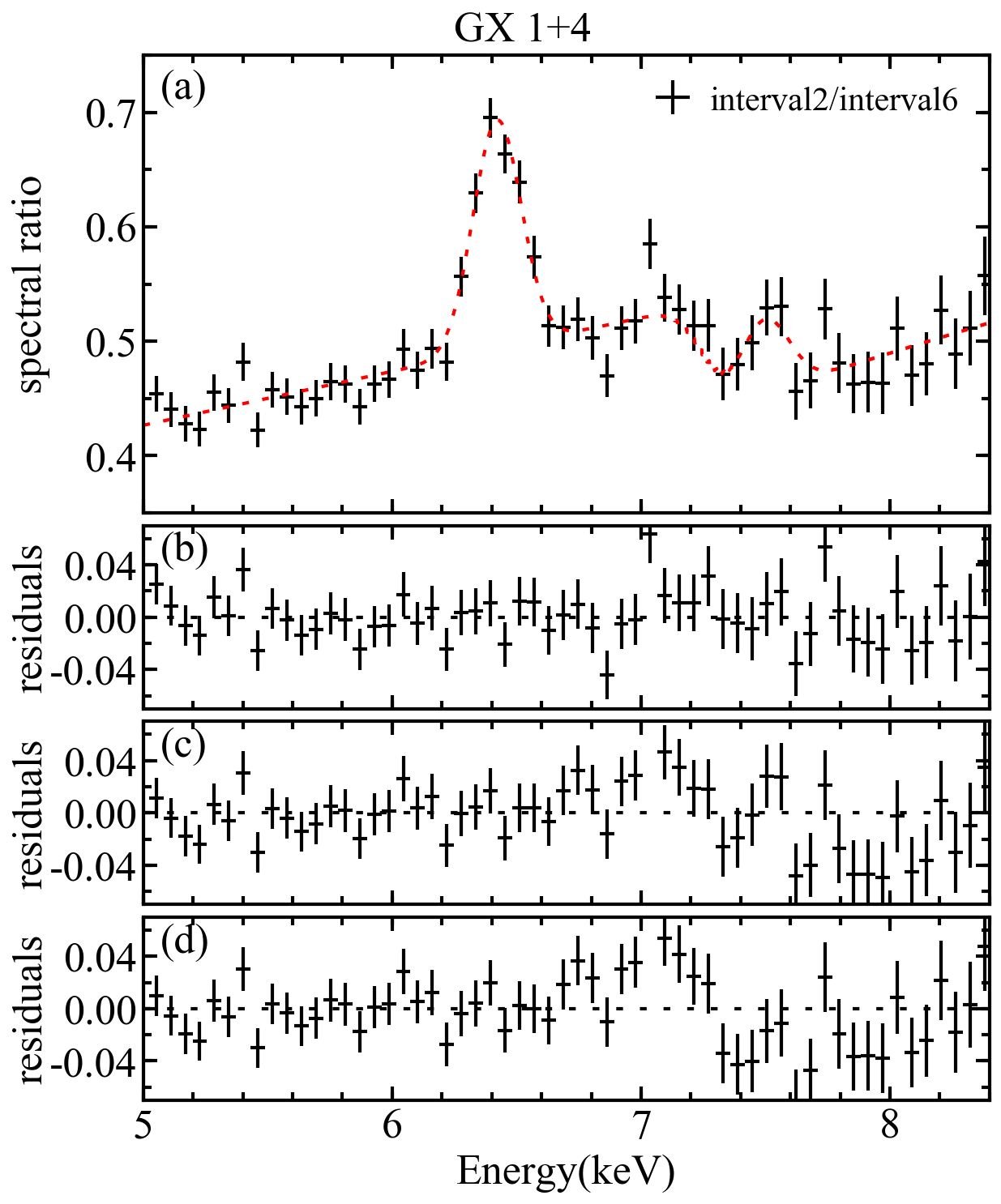}
\caption{As for Figure~\ref{fig:eg_go_sprfit}, but with the results for \gx with data during the whole observation.
\label{fig:eg_gx_sprfit}}
\end{figure}

%%============================================================================================
As noted in the above section, modulations of the optical depth of the iron K-edge with pulse phase can be seen in all four pulsars. 
We performed $\chi^{2}$ tests to evaluate the statistical significance, with a null hypothesis that there is no variation of the optical depth over the pulse phase.
The $\chi^{2}$ values obtained from the constant fittings to the optical depth in the four pulsars are summarized in Table~\ref{tab:sumstattest} with the corresponding null hypothesis probabilities.
These obtained values clearly indicate that the pulse phase modulations in the optical depth are statistically significant. 
To further inspect the significance of the variation in the optical depth, 
we directly compared the spectra extracted from different phase bins by calculating their spectral ratio. 
For clarity, two phase-resolved spectra were chosen so as to maximize the optical depth difference. 
If the optical depths are different and the energies of the edge are the same between the two phases, 
a stepwise feature is expected around the energy of the edge in the spectral ratio. 
The calculated spectral ratio is shown in panel (a) of Figures~\ref{fig:eg_go_sprfit}--\ref{fig:eg_gx_sprfit}.
In the spectral ratios, edge features can be seen around 7.1~keV.
In addition, excesses are exhibited at 6.4~keV and 7.5~keV, 
resulting from variations between the two phases of the equivalent widths of the iron \ka line and nickel \ka line. 

Next, we fitted the spectral ratio in the 5.0--8.4~keV range with the following empirical model:
\begin{eqnarray}
&R(E)=& \left(A \times E+B\right)\times M(E) + G_{\rm Fe}(E) + G_{\rm Ni}(E),\label{eq:edgefit} 
\end{eqnarray}
where $A$ is the slope, $B$ is the intercept, $M(E)$ is the edge component given by Equation~\ref{eq:edge}, and $G_{\rm Fe}(E)$ and $G_{\rm Ni}(E)$ are Gaussian functions expressing the emission line features at 6.4~keV, and 7.5~keV.
In this case, the $\tau_{\rm edge}$ in $ M(E)$ roughly corresponds to the difference in the optical depth at the iron K-edge between the two spectra, denoted by $\Delta$. 
In our fitting, the energy of the edge feature was fixed to the value obtained from the phase-resolved spectral fitting at the phase when the optical depth is maximum. 
The feature of the nickel emission line around 7.5~keV is not clearly separated from the edge feature in the spectral ratio.
Therefore, the ratio of the center energy of $G_{\rm Ni}(E)$ to that of $G_{\rm Fe}(E)$ is fixed at 1.168  for the neutral case,
and the width of $G_{\rm Ni}(E)$ is assumed to be equal to that of $G_{\rm Fe}(E)$. 
The other parameters were allowed to vary. 
In order to simulate smearing due to the detector response of XIS, the model was convolved with a single Gaussian function with a width representing the energy resolution at the observation period. 

The best-fit models are plotted in panel (a) of each figure and residuals from the best-fit model are plotted in panel (b). 
Panel (c) shows residuals from the best-fit model without introducing the edge component, showing stepwise features around 7~keV. 
The depths of the edge feature $\Delta$ derived from the fitting are listed in Table~\ref{tab:sumstattest}. 
It is confirmed that the values of $\Delta$ are consistent with the differences between the optical depths of the two phase-resolved spectra, within the errors, for all sources (see Table~\ref{tab:phasedivfit}). 
The resultant $\chi^{2}$ values are also summarized in Table~\ref{tab:sumstattest}.

To evaluate the statistical significance of the $\chi^{2}$ improvement due to the addition of the edge component, 
we performed an {\it F}-test routine described in \citet{Press:2007aa}.
The {\it F} statistical value is defined as
\begin{eqnarray}
F=\frac{\chi_{1}^{2}/\nu_{1}}{\chi_{2}^{2}/\nu_{2}},
\label{eq:ftest}
\end{eqnarray}
where $\chi_{1}^{2}$ and $\chi_{2}^{2}$ are chi-squared values and $\nu_{1}$ and $\nu_{2}$ are degrees of freedom corresponding to the results of fittings using models~1 and 2 (in this case model~1 is a linear function along with the two line components and model~2 is a linear function multiplied by the edge component with the two line components), respectively. 
The derived {\it F} statistical values are listed in Table~\ref{tab:sumstattest}.
The corresponding chance probabilities in the random case of an improvement of $\chi^{2}$, 
given in Table~\ref{tab:sumstattest}, are not small, and we can only marginally conclude the differences of the optical depths by the  {\it F}-test.

However, the residuals shown in the panel (c) of the above figures clearly show edge features around 7~keV. 
The edge feature is like a stepwise function and might be sensitively tested by a run-test (also called the Wald--Wolfowitz test).
Therefore, we applied the run-test to the residuals obtained from the spectral ratio fitting using the linear function along with the two line components.
We encountered a technical problem in this procedure however, in that we obtained the wrong normalization of $G_{\rm Ni}(E)$ in the fitting without the edge component.
When we fit the data without the edge component, the best-fit continuum component (linear component) became more intense than the data in the energy range above the edge energy and faint below it. 
Even if a hump due to the nickel line exists in the data, the model simulating the nickel line can reproduce only the excess above the continuum model, which should already be stronger than the data. 
Therefore, the obtained best fit normalization of the nickel line by the model without the edge component is smaller than the actual value in the data. 
As a result, the $G_{\rm Ni}(E)$ feature remains in the residuals and prevents a correct evaluation of the edge feature with the run-test.
Therefore, for the run-test, we use the residuals produced by the fitting without the edge component whose normalization of $G_{\rm Ni}(E)$ is fixed at the value obtained from the fit using the edge component. 
The residuals are plotted in panel (d) of each figure. 
The run-test was conducted to evaluate the null hypothesis of the randomness in the residuals in the range 6.5--8.0~keV. 
The number of runs and data points below and above zero are summarized in Table~\ref{tab:sumstattest},
and the derived null hypothesis probabilities from the above run-test are also given.
In the cases of \vl and \gx, the computed probability of random sampling is $<0.01$.
These probabilities suggest a significant detection of the edge-like features in the spectral ratio, and hence the optical depth is indeed different between the two phase bins. 
For \go and \os, similar edge-like features can be seen in the residuals, although they have poor statistical significances from the run-test.

%%%%%%%%%%%%%%%%%%%%%%%%%%%%%%%%%%%%%%%%%%%%%%%%%%%%%%%%%%%%%%
\section{Discussion\label{sec:discussion}}
%%%%%%%%%%%%%%%%%%%%%%%%%%%%%%%%%%%%%%%%%%%%%%%%%%%%%%%%%%%%%%
Focusing on the absorption edges and the emission lines of iron in X-ray pulsars spectra, 
we conducted pulse phase-resolved spectroscopy with data derived from \suzaku observations.
We found a significant modulating optical depth of the iron K-edge with the pulse phase from \vl and \gx.
Similar trends were seen for \go and \os, though with poor statistical significance.
Although the relation between this modulation and the absorption column density derived from the low energy cutoff is interesting, 
we found that the obtained absorption column densities strongly correlate with the power-law index and we consider that the obtained values are not physically meaningful.  
Therefore, this relation is not discussed in this work. 
In the following, we discuss the origin of the absorbing matter based on the parameters obtained for \vl and \gx.
Hereafter we refer to the two phases at which the observed the iron K-edge exhibits the maximum and minimum optical depth as the deepest and shallowest edge phases, respectively.

We assume that the same matter is responsible for the absorption and fluorescent emission, which is illuminated by X-rays from the vicinity of the neutron star. 
Then, we can estimate the ionization state of iron from the ratios of the absorption edge energy to the emission line energy. 
We determined the ionization state to be Fe$_{\rm I\hspace{-0.1em}I\mbox{\textendash}V\hspace{-0.1em}I\hspace{-0.1em}I}$ over the pulse phase.
This ionization state is common between the two pulsars. 
From these ionization states, the ionization parameter $\xi\equiv \frac{L_{\rm X}}{nr^{2}}$ \citep[where $L _{\rm X}$ is the luminosity of the X-ray source, $n$ is the number density of the gas, and $r$ is the distance between the X-ray source and the gas]{Tarter:1969aa} can be estimated. 
The estimated ionization state of iron corresponds to a value of $\xi$ of less than 44.7 ($\log \xi<1.65$) in the case of an optically thick plasma \citep[their model 4, in which the incident X-rays are assumed to be 10-keV bremsstrahlung with $L_{\rm X}=10^{37}$~erg\,s$^{-1}$, roughly simulating the \gx case]{Kallman:1982aa}. 
In the following discussion, we therefore require the ionization parameter of the absorbing matter, so as to satisfy the observed ionization state. 
Unless otherwise specified, the ionization parameter is restricted to $\xi<44.7$.

%%%%%%%%%%%%%%%%%%%%%%%%%%%%%%%%%%%%%%%%%%%%%%%%%%%%%%%%%%%%%%
\subsection{Possible Origin of Modulating Optical Depth of Absorption Edge\label{subsec:model}}
%%%%%%%%%%%%%%%%%%%%%%%%%%%%%%%%%%%%%%%%%%%%%%%%%%%%%%%%%%%%%%
Below, we discuss the following four possible interpretations to explain the modulating optical depth of the absorption edge with the pulse phase:
(1) a variation in the physical state of the absorbing matter, such as the ionization state, according to the spin phase, 
(2) the existence of a modulating component reflected by some neutral matter,
(3) a movement of the emission region, and (4) a change of the absorbing matter location.

%%============================================================================================
\subsubsection{Physical State Variation of Absorbing Matter\label{subsubsec:state}}
%%============================================================================================
The variation in the physical states of the absorbing matter, such as ionization state and density, according to the spin phase, is a possible explanation for the modulating optical depth of the absorption edge with the pulse phase.
Since X-rays from the pulsar are not isotropic, 
the matter located between the observer and the X-ray source is strongly irradiated by the X-rays, 
when the observed X-ray flux of the pulsar is intense. 
The irradiated matter might hence change its ionization state. 
This means that the variation in the optical depth at the absorption edge may result from a change in the number of iron ions in a certain ionization state, 
which are responsible for the absorption edge of interest. 
In this interpretation, the deepest edge phase should correspond to the phase for which the observed X-ray flux from the pulsar decreases, 
while the shallowest edge phase should occur along with an increase of observed X-ray flux from the pulsar.
In addition, an enhancement of the illuminating X-ray intensity to the gas in the line of sight increases the ionization state of the gas,
and the resulting energy of the absorption edge should become high.
On the other hand, the gas recombines as the illuminating X-ray intensity declines, entering a low ionized state, 
and we can expect the energy of the absorption edge to become low.
As a result, this effect can lead to a modulation of the energy with the X-ray flux. 

However, no modulating energy of the edge in association with increasing or decreasing continuous X-ray flux was observed with \suzaku. 
The ratio of the iron K-edge energy to that of the iron \ka line did not change significantly throughout the pulse phase for \vl (see Figure~\ref{fig:eg_vl_paramWphase}).
Furthermore, the energy ratio of \gx becomes large at the deepest edge phase, where the continuous flux becomes minimum (see Figure~\ref{fig:eg_gx_paramWphase}).
This result is the opposite trend than that predicted for the above effect.
These observational results demonstrate that a variation in the physical state of the absorbing matter in the line of sight is not a plausible explanation for the modulating optical depth of the absorption edge with the pulse phase.

%%============================================================================================
\subsubsection{Modulating Reflected X-rays\label{subsubsec:reflect}}
%%============================================================================================
If there is neutral matter surrounding the pulsar, 
a fraction of the X-rays from the neutron star will be reflected by the neutral matter, e.g. a part of the accretion disk, 
and a composite X-ray spectrum will be observed \citep[e.g.,][]{Lightman:1988aa}.
In fact, X-ray reflection on accretion disks is observed in neutron star binaries \citep{Rea:2005aa,Cackett:2010aa}.
The reprocessed X-rays intrinsically exhibit a distinct iron edge feature if the reflecting matter has a composition comparable to the solar composition. 
In addition, a neutral iron emission line coming from the reprocessing matter can be observed. 

We now consider X-ray reflection from neutral matter exposed to X-rays from an X-ray pulsar, whose emission profile is not spherically symmetric. 
In this case, the appearance of the reflected X-ray component can change according to the rotation of the neutron star. 
As a consequence, this modulation of the reflected X-rays can cause variations with the pulse period in the optical depth of the iron edge as well as the flux of the iron emission line of the observed X-ray spectrum.
It is expected that when an increasing X-ray reflection component is observed, 
the optical depth of the edge and the emission line flux will be simultaneously enhanced, in a simple case.

For \gx, although modulations of both the optical depth of the edge and the emission line flux \citep{Yoshida:2017aa} are observed in the \suzaku observation, 
their peak-phases are not synchronized with each other.
In the case of \vl, a variation in the optical depth of the edge is detected with statistical significance, 
while the measured line flux did not change significantly throughout the pulse phase.

However, since there might be other causes changing the emission line flux,
we cannot judge from these observational results whether the modulating reflected X-rays can be attributed to the modulating optical depth of the edge with the pulse phase.

%%============================================================================================
\subsubsection{Movement of the Emission Region\label{subsubsec:emission}}
%%============================================================================================
A change in the amount of matter along the line of sight by a movement of the emission region or of the absorbing matter itself, due to the neutron star spin,
can possibly explain the modulating optical depth of the absorption edge with the pulse phase.  
If the matter is moving with the spin of the neutron star, the matter should exist at or inside the \alfven radius \citep{Lamb:1973aa,Elsner:1977aa}.
This case will be discussed in the next section.

In this section, we suppose that the matter does not move according to the neutron star spin, and hence the absorbing matter exists outside the \alfven radius.
A modulating optical depth with the pulse phase may appear if the emission region moves according to the neutron star spin.
The emission region is thought to exist around the magnetic poles of a neutron star and its size is considered to be less than or comparable to the neutron star radius, namely $10^{6}$~cm. 
On the other hand, the distance of the absorbing matter from the emission region is expected to be more than the \alfven radius, namely $10^{8}$~cm.
By considering this geometry, even if the position of the emission region changes, 
no change of the absorption column density with the pulse phase can be expected, unless there is a very stable and tiny structure in the absorbing matter, which exists outside the \alfven radius.

%%============================================================================================
\subsubsection{Change of Absorbing Matter Location\label{subsubsec:geometry}}
%%============================================================================================
Alternatively, if the absorbing matter co-rotates with the neutron star spin, variation of the optical depth with the pulse phase can occur. 
The co-rotation with the neutron star spin implies that the absorbing matter should be confined by the magnetic field of the neutron star,
and hence it should be located at or within the \alfven radius $R_{\rm A}$.
The outer accretion disk and wind from companion stars cannot be the cause of the pulse phase modulation of the optical depth, 
because these locations cannot have variations with the spin period. 
The \alfven radius is given as 
\begin{eqnarray}
R_{\rm A} &=& 
3.7\times10^{8}\left(M_{\ast}/M_{\odot}\right)^{1/7} R_{6}^{10/7}B_{12}^{4/7}L_{ 37}^{-2/7}\ {\rm cm},
\label{eq:AlfvenRadius}
\end{eqnarray}
where $M_{\ast}$, $R_{6}$, $B_{12}$, and $L_{ 37}$ are the mass, radius (denoted as $R_{\ast}$) in units of $10^{6}$~cm, surface magnetic field strength (denoted as $B_{\rm surf}$) in units of $10^{12}$~G of the neutron star, and luminosity (denoted as $L_{\rm X}$) in units of $10^{37}$~erg\,s$^{-1}$, respectively \citep{Frank:2002aa}.
Substituting a neutron star radius of $R_{\ast}=10$~km and a mass of $M_{\ast}=1.4M_{\odot}$ in Equation~\ref{eq:AlfvenRadius},
\alfven radii for \vl and \gx can be calculated using the estimated X-ray luminosity in the \suzaku observations. 
Here, the magnetic field strength at the neutron star surface of \vl is assumed to be $2.2\times10^{12}$~G, 
derived from the cyclotron resonance scattering feature in the \suzaku X-ray spectrum \citep{Doroshenko:2011aa}. 
However, we tentatively assume that of \gx to be $10^{13}$~G,
since no cyclotron resonance scattering features have been reported in its spectra \citep{Yoshida:2017aa}.

The absorbing matter responsible for the absorption edge should be located within the \alfven radius.
This fact puts restrictions on the distance between the X-ray source and the matter, $r$, as 
\begin{eqnarray}
r\leqslant 8.1\times10^{8}\,{\rm cm}
\label{eq:requirement1_vl}
\end{eqnarray}
for \vl ($L_{\rm X}=3.1\times10^{36}\ {\rm erg\,s^{-1}}$ and $B_{\rm surf}=2.2\times10^{12}\,{\rm G}$) and
\begin{eqnarray}
r\leqslant 1.4\times10^{9}\,{\rm cm}
\label{eq:requirement1_gx}
\end{eqnarray}
for \gx ($L_{\rm X}=8.9\times10^{36}\ {\rm erg\,s^{-1}}$ and $B_{\rm surf}=1.0\times10^{13}\,{\rm G}$).
The absorbing matter should also satisfy the constraint due to the ionization parameter based on the observed ionization state of iron.
Specifically, the restrictions are written as 
\begin{eqnarray}
nr^{2}>6.9\times10^{34}~{\rm cm}^{-1},
\label{eq:requirement2_vl}
\end{eqnarray}
and
\begin{eqnarray}
nr^{2}>2.0\times10^{35}~{\rm cm}^{-1}
\label{eq:requirement2_gx}
\end{eqnarray}
for \vl and \gx, respectively, where $n$ is the number density of particles of the matter.
The acceptable regions on the $r$--$n$ planes satisfying the requirement on the absorbing matter given by the above equations, 
 \ref{eq:requirement1_vl}, \ref{eq:requirement1_gx}, \ref{eq:requirement2_vl}, and \ref{eq:requirement2_gx}, are shown in light gray in panels (a) and (b) of Figure~\ref{fig:rndiagram}, for \vl and \gx, respectively. 
For the observed amount of change over the pulse phase in the hydrogen column density of $\Delta N_{\rm H}=10^{23}$~cm$^{-2}$, 
the geometric thickness of the absorbing matter along the line of sight, $\delta$, can be calculated for a given particle density, and is indicated by the vertical axes on the right of the panels of Figure~\ref{fig:rndiagram}. 
If the absorbing matter is at the \alfven radius and has a particle density of $n=10^{18}$~${\rm cm}^{-3}$, 
the absorbing matter contains almost neutral irons and forms a structure whose geometric thickness along the line of sight is $10^{5}$~cm.
Given these facts, the observed variation in the amount of absorption in the line of sight due to the absorbing matter co-rotating with the neutron star spin is a plausible explanation for the modulating optical depth of the iron absorption edge with the pulse phase.

%%%%%%%%%%%%%%%%%%%%%%%%%%%%%%%%%%%%%%%%%%%%%%%%%%%%%%%%%%%%%%
\subsection{Accretion Flow as Absorbing Matter\label{subsec:flow}}
\begin{figure*}[ht]
\centering
\includegraphics[bb= 0 0 560 554,width=0.48\textwidth]{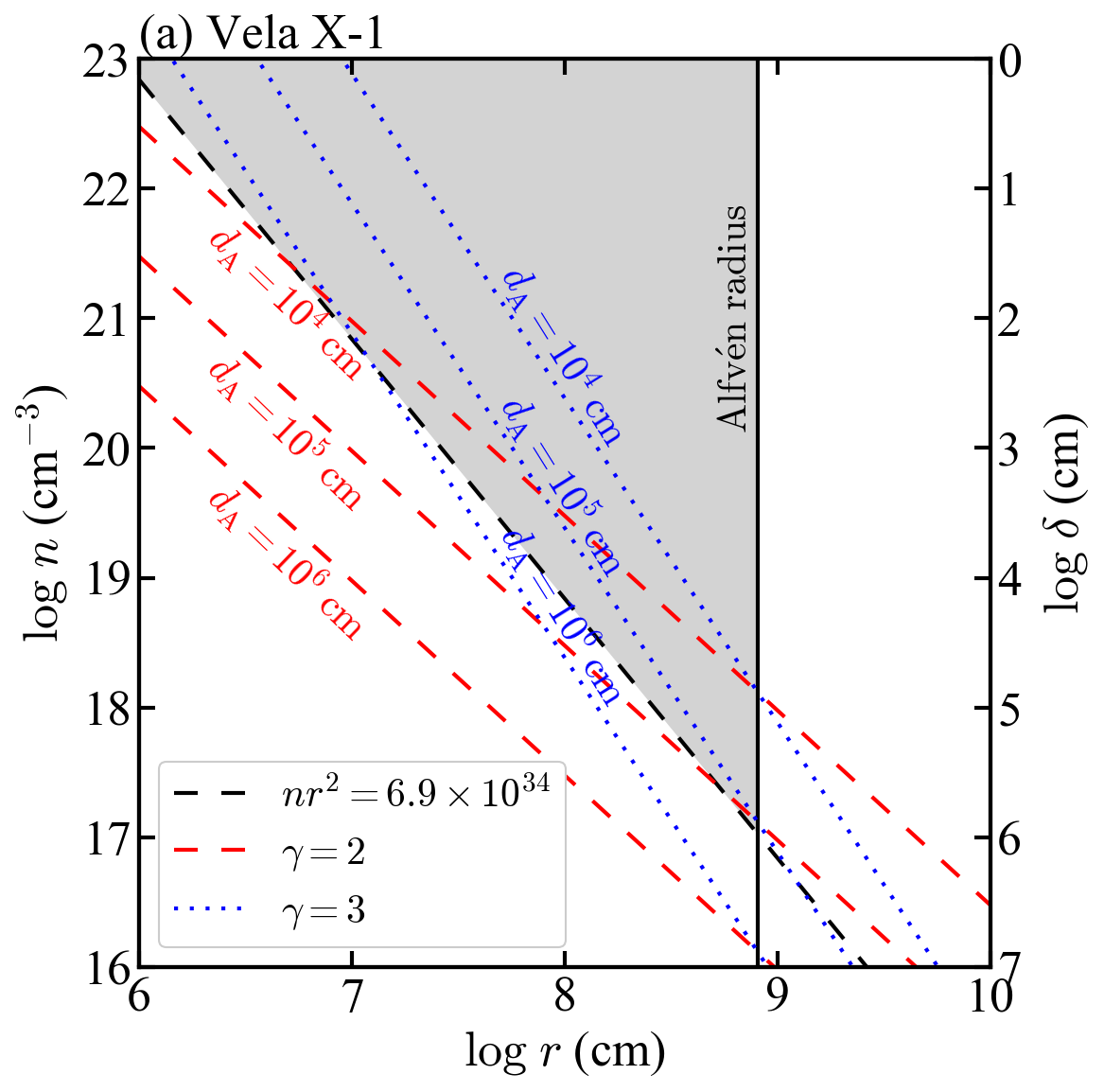}
\includegraphics[bb= 0 0 560 554,width=0.48\textwidth]{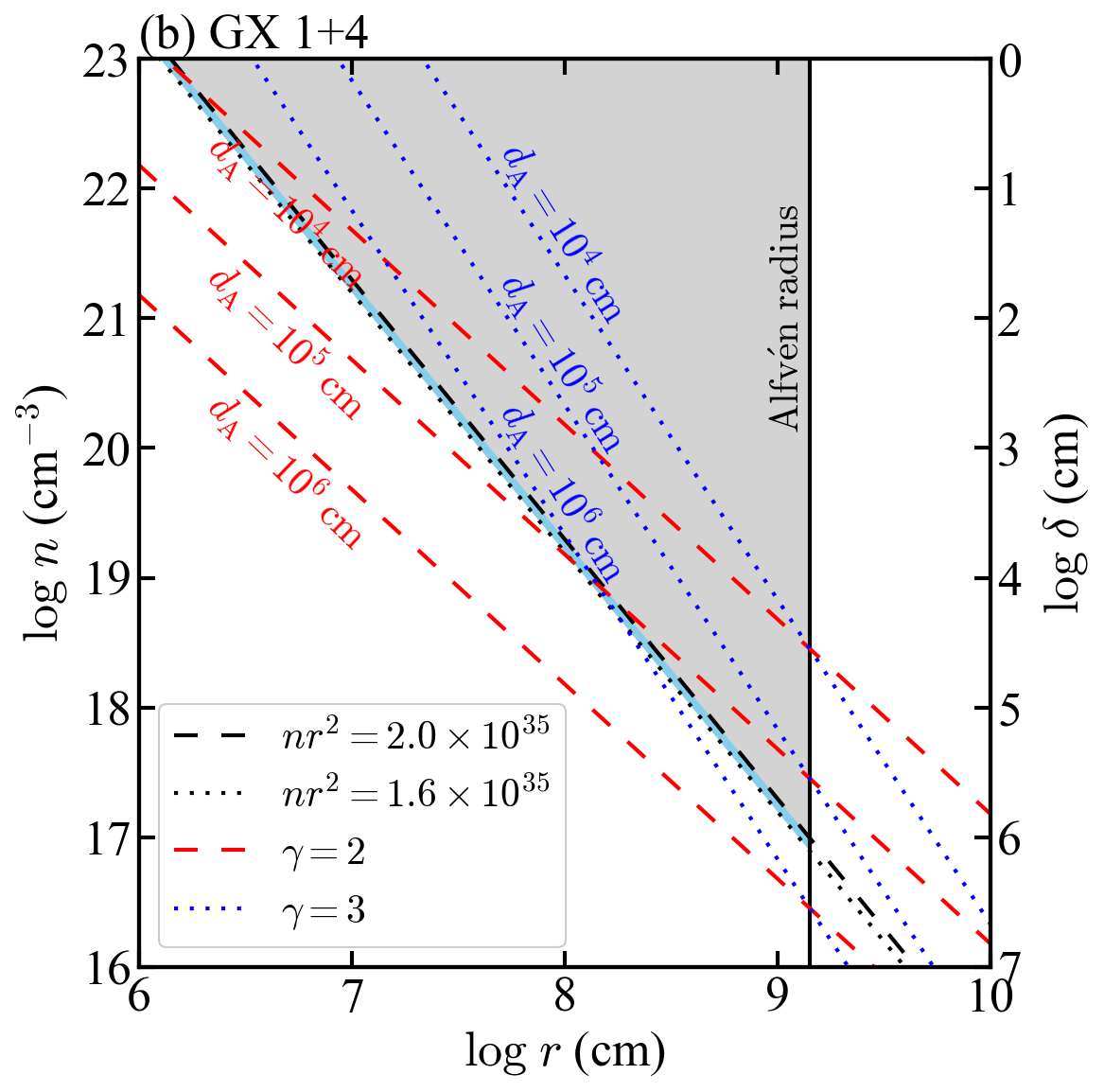}
 \caption{
Constraints on the absorbing matter for (a) \vl and (b) \gx, indicated on the diagram between the number density of particles and the distance from the neutron star ($r$--$n$ plane).
The black solid lines in each panel indicate the \alfven radii for each source. 
The black dashed lines in each panel correspond to the boundaries of Equation~\ref{eq:requirement2_vl} and \ref{eq:requirement2_gx},
and black dotted line in panel (b) corresponds to the boundary of Equation~\ref{eq:requirement3_gx}.
The light gray colored regions indicate the allowed regions for both restrictions of Equation~\ref{eq:requirement1_vl} and \ref{eq:requirement2_vl} in panel~(a), and those for both restrictions of Equation~\ref{eq:requirement1_gx} and \ref{eq:requirement2_gx} in panel~(b),
while the light blue colored region in panel~(b) satisfies the requirements of both Equation~\ref{eq:requirement1_gx} and \ref{eq:requirement3_gx}.
The vertical axes on the right of each panel indicate the expected thickness of absorbing matter along the line of sight.
The estimated densities in the accretion flow given by Equation~\ref{eq:nincolumn} are indicated by red-dashed and blue-dotted lines for $\gamma=2$ and 3, respectively. 
See text for details. 
 \label{fig:rndiagram}}
\end{figure*}

\begin{figure}[ht]
\centering
\includegraphics[bb= 0 0 145 227,width=0.3\textwidth]{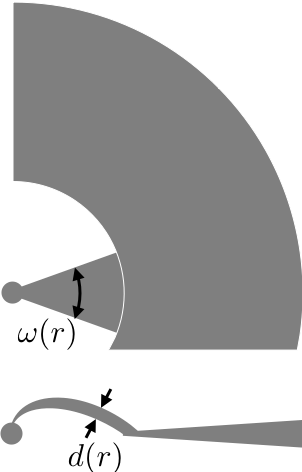}
 \caption{
Schematic picture of the accretion geometry. 
The top figure shows the top view and the bottom figure shows the sectional view.
 \label{fig:accretionflow} }
\end{figure}

%%%%%%%%%%%%%%%%%%%%%%%%%%%%%%%%%%%%%%%%%%%%%%%%%%%%%%%%%%%%%%
The condition described in \S~\ref{subsubsec:geometry} suggests that 
the absorbing matter is the matter accreting onto the neutron star along its magnetic field lines.
We can therefore discuss the geometry of the accretion flow 
and determine the relation between the mass accretion rate $\dot{M}$, $r$, $n(r)$, and the falling velocity $v(r)$.
We thus obtain
\begin{eqnarray}
\dot{M}&=&{\mu_{\rm gas} m_{\rm p}n(r)s(r)v(r)} \nonumber\\
&=&{\sqrt{2GM_{\ast}/r}\mu_{\rm gas}m_{\rm p}n(r)s(r)},
\label{eq:massconservation}
\end{eqnarray}
where $s(r)$ is the cross sectional area of the accretion flow, $m_{\rm p}$ is the proton mass, 
and $\mu_{\rm gas}$ is the mean atomic mass of the cosmic material, which is 1.3 for solar abundances \citep{Cox:2000aa}. 
We assumed the velocity falling onto the neutron star to be the free fall velocity given by $v(r)= \sqrt{2GM_{\ast}/r}$. 
By assuming that all the kinetic energy liberated is converted to radiation energy at the neutron star surface, 
namely $L_{\rm X}=GM_{\ast}\dot{M}/R_{\ast}$, the density in the accretion flow is given by
\begin{eqnarray}
n(r)&=&2.5\times10^{27} r^{1/2} s(r)^{-1}\nonumber \\ 
&&\times\left(M_{\ast}/M_{\odot}\right)^{-3/2}R_{6}L_{37}\ {\rm cm}^{-3}.
\label{eq:nincolumn}
\end{eqnarray}
The cross sectional area of the accretion flow is given by the product of the width and thickness of the accretion flow, $w(r)$ and $d(r)$,
and is assumed to be proportional to $r^{\gamma}$. 
A schematic picture of the assumed accretion flow is shown in Figure~\ref{fig:accretionflow}. 
We assume that the width is $w(r)=\zeta r$, where $\zeta$ is a constant azimuthal angle of the accretion flow,  
and that the thickness can be written as $d(r)=d_{\rm A}\left(r/R_{\rm A}\right)^{\gamma-1}$, where $d_{\rm A}$ is the thickness at $R_{\rm A}$.
Then, we obtain
\begin{eqnarray}
s(r) = \zeta d_{\rm A}{R_{\rm A}}^{1-\gamma} {r}^{\gamma}.
\label{eq:columncrosssection}
\end{eqnarray}
In the case of \gx, the azimuthal angle of the accretion flow can be estimated from the pulse phase duration during the deepest edge phase, 
yielding $\zeta=0.4\pi$ (see Figure~\ref{fig:eg_gx_paramWphase}).
We also assume the same condition for \vl, although we have no basis for this assumption. 
If we consider a simple conic geometry for the accretion flow, the index of $\gamma$ is 2.
On the other hand, \citet{Ghosh:1979aa} reported that the cross sectional area of the accretion flow $s(r)$ can be assumed to be proportional to the cube of $r$, namely $\gamma=3$.
Since this assumption of $\gamma=3$ can be applied only near the neutron star and cannot be extrapolated to the \alfven radius,
we consider that the actual $\gamma$ value should be between 2 and 3. 
Thus, we calculated both cases of $\gamma=2$ and 3.
Substituting Equation~\ref{eq:columncrosssection} into Equation~\ref{eq:nincolumn} with $\gamma=2$ and 3,
we calculated the number density in the accretion flow. 
Using the obtained parameters for \vl and \gx, i.e., the X-ray luminosities, magnetic strengths at the neutron star surface, and $\zeta=0.4\pi$, 
the calculated number densities in the accretion flow are plotted by red dashed lines ($\gamma=2$) and blue dotted lines ($\gamma=3$) in Figure~\ref{fig:rndiagram} for some typical thicknesses for the accretion flow at the \alfven radius,
$d_{\rm A}=10^{4}$, $10^{5}$, and $10^{6}$~cm. 
For both \vl and \gx, as seen in Figure~\ref{fig:rndiagram}, the accreting matter at any radius is acceptable
if it has a thickness of $d_{\rm A}<10^{5}$~cm and its cross sectional area is proportional to $r^{3}$.
On the other hand, for $\gamma=2$, only the outer region is acceptable for accreting matter of thickness $d_{\rm A}\sim10^{4\mbox{\textendash}5}$~cm, for both sources.
It should be noted that the thickness of the modeled accretion flow at the \alfven radius, $d_{\rm A}$, is roughly consistent, within a factor of 3--8, with $\delta$, indicated on the vertical axes on the right of the panels in Figure~\ref{fig:rndiagram}. 

As mentioned in \S~\ref{subsubsec:spectroscopy}, the deepest edge phase corresponds to the dimming interval of the X-ray continuum, such as a dip interval. 
The dip in the pulse profile is interpreted as being due to the eclipse of the X-ray emitting region by the accretion column of the pulsar, at least for \gx \citep{Dotani:1989aa,Giles:2000aa,Galloway:2000ab,Galloway:2001aa}.
\citet{Galloway:2001aa} reported that the absorption column density, derived from the low energy cutoff, increased at the dip.
This is in good agreement with the results described here. 
The increasing amount of absorption could be attributed to a crossing of the line of sight by the accretion column or curtain \citep{Miller:1996aa}. 
In our geometry of the accretion flow with Equation~\ref{eq:columncrosssection}, 
its thickness is sufficiently smaller than its width
if the thickness is the estimated value of $10^{4{\mbox \textendash}6}$~cm at the \alfven radius.
Therefore, for the configuration of the accretion flow, 
a geometry that is flattened along the azimuthal direction like an accretion curtain \citep[latter shows the schematic figure of accretion geometry]{Miller:1996aa,Galloway:2001aa} may be more plausible than a whole cone geometry.

%%%%%%%%%%%%%%%%%%%%%%%%%%%%%%%%%%%%%%%%%%%%%%%%%%%%%%%%%%%%%%
\subsection{Distribution of Iron Surrounding X-ray Pulsar\label{subsec:distribution}}
%%%%%%%%%%%%%%%%%%%%%%%%%%%%%%%%%%%%%%%%%%%%%%%%%%%%%%%%%%%%%%
The observed modulating optical depth of the iron K-edge with pulse period can be explained by the absorption by the accreting matter along the magnetic field lines within the \alfven radius, which co-rotates with the neutron star.
That is to say, the shallowest edge phase can be interpreted as the phase when the accreting matter is out of the line of sight.
However, we note that even at the shallowest edge phase, 
a large amount of absorption, corresponding to $N_{\rm H}=10^{23}$~cm$^{-2}$, was observed (see Table~\ref{tab:phasedivfit}).  
We cannot determine whether this large absorption is caused by the accreting matter within the \alfven radius 
or by the circumstellar matter located outside the \alfven radius.

Several authors have discussed the possibility that a reprocessing site contributes to the 6.4~keV iron fluorescence line emission. 
For \vl, the presence of a line emission site within $5\times10^{11}$~cm of the neutron star was proposed as an explanation for a drop of the iron line flux of a factor of 20 during the eclipse observed by {\it Ginga} \citep{Ohashi:1984aa}. 
For \gx, it was reported that the peak-phase of the modulating line flux is delayed compared to that of the pulse profile of the continuum by $\sim$30~s \citep{Yoshida:2017aa}.
A possible explanation for this delay was proposed by \citet{Yoshida:2017ab}, where the iron line is a fluorescence of the matter exposed to the X-ray continuum from the neutron star, at a distance of $\sim10^{12}$~cm from the X-ray source. 
Therefore, it is natural that a certain amount of matter exists outside the \alfven radius.
If so, the matter located in this region should be responsible for not only the fluorescent but also the photoelectric absorption. 
The absorption contribution by the matter in this region is not affected by the spin of neutron stars and is a possible candidate for the absorption observed at the shallowest edge phase.

%%%%%%%%%%%%%%%%%%%%%%%%%%%%%%%%%%%%%%%%%%%%%%%%%%%%%%%%%%%%%%
\subsection{Two Absorption Edges Hypothesis\label{subsec:two}}
%%%%%%%%%%%%%%%%%%%%%%%%%%%%%%%%%%%%%%%%%%%%%%%%%%%%%%%%%%%%%%
In the above section, we argued that the absorbing matter may be distributed in two regions separated by the \alfven radius. 
The matter within the \alfven radius has an asymmetric structure and is the origin of the modulating absorption,
while the matter outside the \alfven radius is responsible for the phase independent absorption. 
In our analysis described in \S~\ref{subsec:resolve}, we assumed that the absorbing matter is one component, 
and that there is no variety in the ionization state of the absorbing matter along the line of sight. 
We then determined the ionization state of iron from the ratio between the energies of the iron \ka emission line and the iron absorption K-edge. 
However, the ionization states of the absorbing matter in the two regions does not need to be the same. 
The matter causing the modulation in the absorption and the other matter causing constant absorption should be handled separately in the analysis. 
Therefore, we conducted an additional spectral analysis dealing with the absorption by the matter in the two regions separately. 

\begin{figure*}[ht]
\centering
\includegraphics[bb= 0 0 717 601,width=0.48\textwidth]{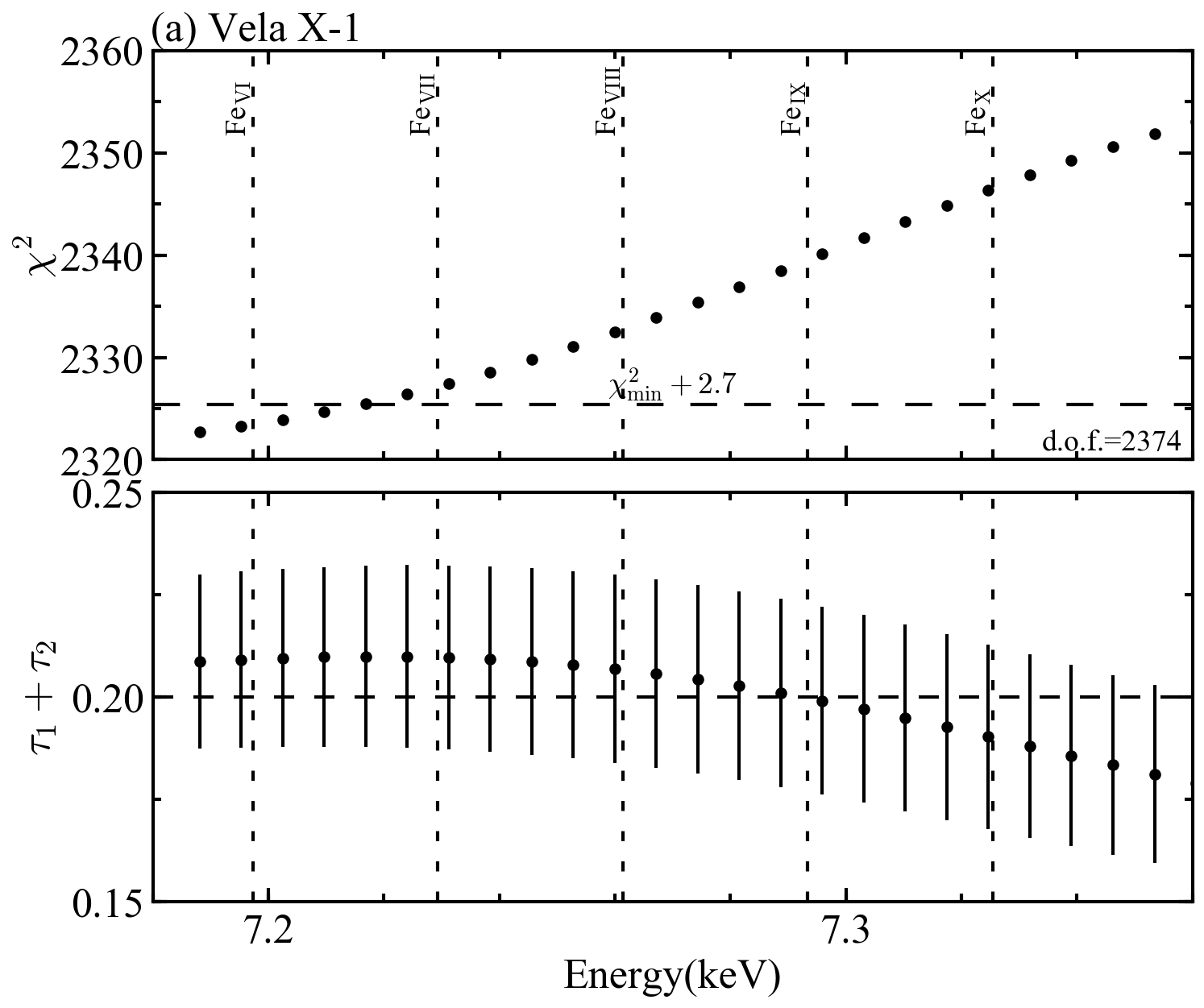}
\includegraphics[bb= 0 0 717 601,width=0.48\textwidth]{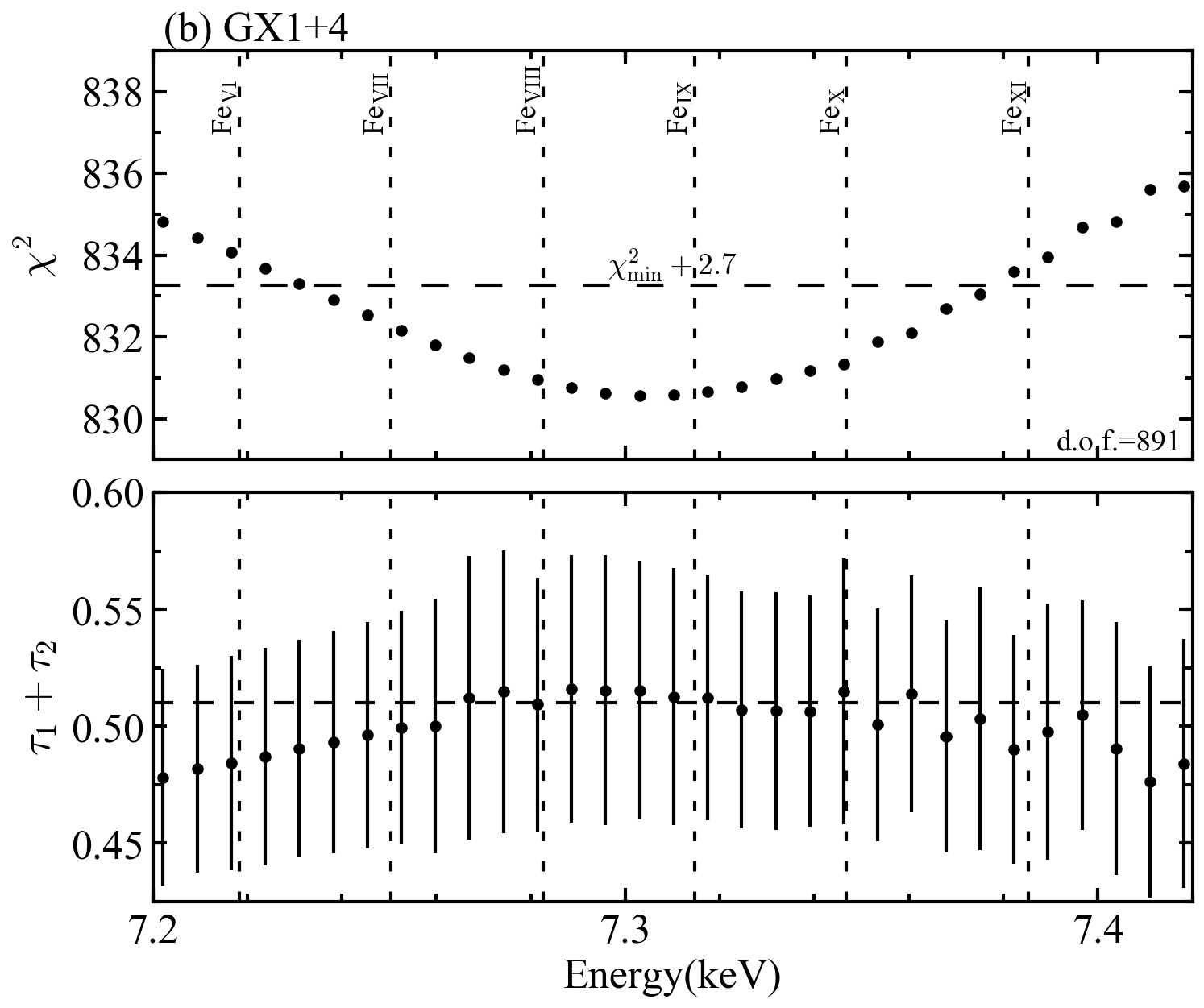}
\caption{
Upper panels: resultant $\chi^{2}$ values as a function of energy of the iron K-edge (${\tt edge}_{2}$), obtained from the fitting with two absorption edge components with degrees of freedom of 2374 for \vl and 891 for \gx. 
The horizontal dashed lines represent the 90\% confidence level.
The expected energies of the iron K-edge for some cases of the ionization state are indicated by vertical dashed lines.
Lower panels: sum of optical depths of the two absorption edge components. 
The horizontal dashed lines indicate the optical depth at the deepest edge phase obtained with the single absorption edge component. \label{fig:eg_eEdgeWchi2}}
\end{figure*}

We fitted the phase-resolved spectra of \vl and \gx at the deepest edge phase, 
restricting the energy range to 5.0--7.9~keV, similar to that in \S~\ref{subsec:resolve}. 
For the spectral fitting in the restricted energy range, 
a model consisting of ${\tt powerlaw}\times{\tt edge_{1}}\times{\tt edge_{2}}+{\tt gaussian}+{\tt gaussian}+{\tt gaussian}$ is applied to express the absorption by iron in the two regions. 
In this model, ${\tt edge_{1}}$ and ${\tt edge_{2}}$ represent the absorption edges of the matter outside and within the \alfven radius, respectively. 
In the fitting, the ratio of the energy of the iron \kb line to that of the iron \ka line was fixed to 1.103 for the neutral case \citep{Yamaguchi:2014aa}.
The center energy of the nickel \ka line was also fixed to a value derived from the phase-resolved fitting at the deepest edge phase. 
The energy and the optical depth of the ${\tt edge_{1}}$ component were fixed to the values obtained from the fitting of the shallowest edge phase.
The energy of the ${\tt edge_{2}}$ component was scanned by changing the ratio to the fixed edge energy from 1.00 to 1.05 with steps of 0.001,
while the optical depth of the ${\tt edge_{2}}$ component was allowed to be free.
The resultant $\chi^{2}$ values as a function of the energy of the ${\tt edge_{2}}$ are plotted in the upper panels of Figure~\ref{fig:eg_eEdgeWchi2}(a) and (b), for \vl and \gx.
In the panels, the 90\% confidence levels are indicated by horizontal dashed lines.
In addition, the energies of the iron K-edge for several cases of the ionization state are expressed by vertical dashed lines,
calculated from the energy of the iron \ka line at the deepest edge phase and the energy ratio between the absorption edge and the \ka line of iron \citep{Kallman:2004aa,Yamaguchi:2014aa}.
The energy of ${\tt edge_{2}}$ are constrained to 7.23--7.40~keV at the 90\% confidence level for \gx, 
while for \vl, an upper limit of 7.22~keV at the 90\% confidence level is determined.
These energy ranges correspond to the ionization states of Fe$_{\rm V\hspace{-0.1em}I{\mbox \textendash}X\hspace{-0.1em}I}$,
and $<$Fe$_{\rm V\hspace{-0.1em}I\hspace{-0.1em}I}$, respectively.
The determined ionization state of iron within the \alfven radius for \gx is equal to or higher than that obtained from the phase-resolved fitting with the assumption of a single absorption edge component, 
while in the case of \vl, it is the same ionization state as in the phase-resolved analysis. 
The lower panels of the figure show the sum of the optical depths of the two edge components, $\tau_{1}+\tau_{2}$, as a function of the energy of ${\tt edge_{2}}$.
As long as the energy of the ${\tt edge_{2}}$ component is in an acceptable range, 
these values are consistent within the errors with the maximum depths derived from the phase-resolved fitting employing the single absorption edge component, 
indicated by horizontal dashed lines in the panels.

Now, we again discuss the physical properties, number density, and distance from the neutron star, of the absorbing matter existing within the \alfven radius with the ionization state obtained from the above additional analysis. 
The determined restrictions of the ionization state of iron, Fe$_{\rm V\hspace{-0.1em}I{\mbox \textendash}X\hspace{-0.1em}I}$, for \gx roughly correspond to the value of the ionization parameter $44.7<\xi<56.2$ ($1.65<\log \xi<1.75$) in the case of an optically thick plasma \citep[their model 4]{Kallman:1982aa}.
The absorbing matter within the \alfven radius therefore should satisfy the following conditions:
\begin{eqnarray}
1.6\times10^{35}~{\rm cm}^{-1}<nr^{2}<2.0\times10^{35}~{\rm cm}^{-1}
\label{eq:requirement3_gx}
\end{eqnarray}
for \gx, so as to satisfy the observed ionization state. 
Based on this condition, combined with the requirement on the radius (Equation~\ref{eq:requirement1_gx}),
the acceptable region for the absorbing matter within the \alfven radius can be represented on the $r$--$n$ plane.
It is indicated by the light blue region in panel (b) of Figure~\ref{fig:rndiagram}. 

If the accreting matter along the magnetic field lines at the \alfven radius has a particle density of $n=10^{17}$~cm$^{-3}$, the ionization state of iron can be Fe$_{\rm V\hspace{-0.1em}I{\mbox \textendash}X\hspace{-0.1em}I}$ and the thickness of the accretion flow is $10^{5{\mbox \textendash}6}$~cm.   
This accretion matter can cause the modulating optical depth of the absorption edge with the spin period.

If the modulating optical depth of the absorption edge with the pulse phase originates in the asymmetric structure of the accreting matter co-rotating with the pulsar spin, 
a variation of the ionization state of the absorbing matter along the line of sight, namely a modulation of the energy of the edge, can be expected.   
Also, a possible modulation of the center energy of the iron \ka line, caused by the asymmetric accreting matter, is expected due to the variation of the flow velocity along the line of sight. 
Significantly improved sensitivity and energy resolution from future X-ray missions, such as {\it XRISM}, will allow further detailed observations of the absorption edge, as well as of the emission lines.

%%%%%%%%%%%%%%%%%%%%%%%%%%%%%%%%%%%%%%%%%%%%%%%%%%%%%%%%%%%%%%
\section{Conclusion\label{sec:conclusion}}
%%%%%%%%%%%%%%%%%%%%%%%%%%%%%%%%%%%%%%%%%%%%%%%%%%%%%%%%%%%%%%
We performed a study of X-ray pulsars focusing on the absorption edge and the emission line of iron,
and investigated the properties of the iron surrounding the pulsar and the accretion flow along the magnetic field. 

We discovered a significantly modulating optical depth of the iron K-edge with the neutron star spin for \vl and \gx.
Similar trends were found for \go and \os, though with poor statistical significance. 
A possible interpretation of the observed changes of the optical depth with pulse phase is that the accreting matter captured by the magnetic field lines of the pulsar, which co-rotates with the neutron star spin, causes photoelectric absorption. 
Based on the observations of the iron absorption edge, 
we speculate that iron surrounding an X-ray pulsar is distributed in two regions divided by the \alfven radius,
where the iron is in different physical conditions of ionization state and geometry.   
In particular, we propose that the accretion flow along the magnetic field line of a neutron star within the \alfven radius contains iron with an ionization state of Fe$_{\rm V\hspace{-0.1em}I{\mbox \textendash}X\hspace{-0.1em}I}$, 
with a particle density of $n\sim10^{17}$~cm$^{-3}$.

%%%%%%%%%%%%%%%%%%%%%%%%%%%%%%%%%%%%%%%%%%%%%%%%%%%%%%%%%%%%%%
\acknowledgments
%%%%%%%%%%%%%%%%%%%%%%%%%%%%%%%%%%%%%%%%%%%%%%%%%%%%%%%%%%%%%%
This research was carried out using data obtained from the Data Archive and Transmission System (DARTS), 
provided by the Center for Science-satellite Operation and Data Archive (C-SODA) at ISAS/JAXA.
This work was partially supported by a MEXT-Supported Program for the Strategic Research Foundation at Private Universities, 2014--2018 (S1411024).

%%%%%%%%%%%%%%%%%%%%%%%%%%%%%%%%%%%%%%%%%%%%%%%%%%%%%%%%%%%%%%
%% TABELS
%%%%%%%%%%%%%%%%%%%%%%%%%%%%%%%%%%%%%%%%%%%%%%%%%%%%%%%%%%%%%%

%% TABEL of Observation Log
\begin{table*}[ht!]
\begin{center}
\caption{Log of the \suzaku observations of the four X-ray pulsars.
\label{tab:obs_lst}}
\begin{tabular}{lllDD@{\hspace{0.1em}/}D@{\hspace{0.1em}/}Dcl}
\tablewidth{0pt}
\tableline\tableline
Source\tablenotemark{a}	&ObsID	& UT start/end time		& \multicolumn{2}{l}{Duration\tablenotemark{b}}		& \multicolumn{6}{l}{Exposure\tablenotemark{c}}	& Aim\tablenotemark{d}	& XIS Operation\tablenotemark{e}  \\
					&		& 					& \multicolumn{2}{l}{}							& \multicolumn{6}{l}{XIS\,0\,/\,XIS\,1\,/\,XIS\,3}		&					& clock,window \\
					&		& (YYYY-MM-DD hh:mm)	& \multicolumn{2}{l}{(ks)}						& \multicolumn{6}{l}{(ks)}						&					&		  \\
\tableline
\decimals
%\go	& 403044010	& 2008-08-25 13:15/2008-08-26 00:05	& 39.0    	& 11.4    	& 11.4    	& 11.4    	& HXD	& normal,1/4 \\
\go	& 403044020	& 2009-01-05 10:33/2009-01-07 01:00	& 138.4    	& 61.8    	& 61.8    	& 61.8    	& HXD	& normal,1/4 \\
\vl	& 403045010	& 2008-06-17 04:45/2008-06-18 21:42	& 147.4    	& 104.7   	& 104.7    	& 104.7    	& XIS	& normal,1/4 \\
\os	& 406011010	& 2011-09-26 09:34/2011-09-28 16:00	& 195.9    	& 84.8     	& 84.8    	& 84.8    	& XIS	& normal,1/4 \\
\gx	& 405077010	& 2010-10-02 06:43/2010-10-04 12:20	& 193.0  	& 97.3  	& 99.7    	& 88.3    	& HXD	& normal,1/4 \\
\tableline
\end{tabular}
\end{center}
%\tablecomments{}
\tablenotetext{a}{\suzaku studies: \go \citep{Suchy:2012aa}, \vl \citep{Doroshenko:2011aa,Odaka:2013aa,Maitra:2013aa}, \os \citep{Pradhan:2014aa,Jaisawal:2014aa}, \gx \citep{Yoshida:2017aa}}
\tablenotetext{b}{Observational duration time.}
\tablenotetext{c}{Effective exposure times with each XIS.}
\tablenotetext{d}{Nominal pointing position (XIS or HXD) in the observations \citep[see][]{Mitsuda:2007aa}.}
\tablenotetext{e}{Clock mode and window option of XIS \citep[see][]{Koyama:2007aa}.}
\end{table*}

%% TABEL of Pulse Period
\begin{table*}[ht!]
\begin{center}
\caption{Results of time-averaged analysis of the four X-ray pulsars from \suzaku observations.
\label{tab:TimAveResult}}
\begin{tabular}{lDlllllll@{\,/\,}l}
\tablewidth{0pt}
\tableline\tableline
Source & P{\rm (s)}\tablenotemark{a} & Epoch\tablenotemark{b} & Model\tablenotemark{c} & $N_{\rm H1}$\tablenotemark{d} & $Z_{\rm Fe}$\tablenotemark{e} & $N_{\rm H2}$\tablenotemark{f} & $f_{\rm pc}$\tablenotemark{g} & \multicolumn{2}{l}{$L_{\rm X}\tablenotemark{h}\,/\,D$\tablenotemark{i}}    \\
\tableline
\decimals
\go & 685.532(4) & 54836.44 & FDPL & $31\pm2$ & 1 (fixed) & $33^{+2}_{-1}$ & $0.77\pm0.05$ & 8.5 &3.0 \\
\vl & 283.449(2) & 54634.22 & NPEX & $0.77^{+0.08}_{-0.07}$ & $12\pm2$\tablenotemark{j} & $8.9\pm0.2$ & $0.466^{+0.007}_{-0.008}$ & 3.1&1.9 \\
\os & 36.96(1) & 55832.06 & NPEX & $25.8\pm0.9$ & 1 (fixed) & $57^{+6}_{-5}$ & $0.59^{+0.03}_{-0.02}$ & 20 & 7.1 \\
\gx & 159.944(1) & 55471.28 & BB+CPL & $18.2^{+0.3}_{-0.4}$ & $1.43\pm0.08$ & \mbox{$\cdots$} & \mbox{$\cdots$} & 8.9 &4.3 \\
\tableline
\end{tabular}
\end{center}
%\tablecomments{}
\tablenotetext{a}{Barycentric pulsation period  obtained from HXD-PIN data in the range 12.0--70.0~keV. The quoted errors are the statistical 1$\sigma$ level.}
\tablenotetext{b}{Epoch to fold the light curves with individual pulse period (MJD).}
\tablenotetext{c}{Applied continuum model to fit the X-ray broad-band spectra. 
BB, CPL, NPEX, and FDPL represent blackbody, exponential cutoff power-law, negative and positive power-law with an exponential cutoff, and power-law multiplied Fermi--Dirac cutoff, respectively.} 
\tablenotetext{d}{Equivalent hydrogen column densities ($10^{22}$\,H\,atoms~cm$^{-2}$) estimated from the fully covering photoelectric absorption model.}
\tablenotetext{e}{Iron abundance derived from the photoelectric absorption model.}
\tablenotetext{f}{Equivalent hydrogen column densities ($10^{22}$\,H\,atoms~cm$^{-2}$) estimated from the partially covering photoelectric absorption model.}
\tablenotetext{g}{Covering fraction of the partial covering photoelectric absorption model. }
\tablenotetext{h}{X-ray luminosity ($10^{36}$~erg\,s$^{-1}$) in the range 0.5--100~keV calculated with a distance of $D$. Ambiguities of the distance are not included.}
\tablenotetext{i}{Distance to  the source from the observer (kpc), from \citet{Kaper:2006aa,Sadakane:1985aa,Audley:2006aa,Hinkle:2006aa} for \go, \vl, \os, and \gx, respectively.}
\tablenotetext{j}{This value is thought to be wrong. See text for details.}
\end{table*}

%% TABEL of Results of Phase-Resolved Spectral Fitting
\begin{table*}[ht]
\begin{center}
\caption{Results of phase-resolved spectral fitting.
\label{tab:phasedivfit}}
\begin{tabular}{clllllll}
\tablewidth{0pt}
\tableline\tableline
Interval\tablenotemark{a} & $E_{\rm FeK_{\alpha}}$\tablenotemark{b}& $E_{\rm FeK_{edge}}$\tablenotemark{b} & $\tau_{\rm edge}$\tablenotemark{c} & $N_{\rm H}$\tablenotemark{d} &$I_{\rm FeK_{\alpha}}$\tablenotemark{e} & $I_{\rm FeK_{\beta}}$\tablenotemark{e} & $\chi^{2}_{\nu}$\,(d.o.f.) \\
\tableline
%
%%GX301-2
\multicolumn{8}{c}{\go}\\
1&$6.392\pm0.004$&$7.152^{+0.007}_{-0.01}$&$0.68^{+0.02}_{-0.04}$&6.4&$2.04^{+0.09}_{-0.07}$&$<0.08$&0.98\,(1663)\\
2&$6.392\pm0.002$&$7.142^{+0.007}_{-0.01}$&$0.72\pm0.03$&6.8&$1.97\pm0.05$&$<0.08$&1.09\,(2126)\\
3&$6.392^{+0.004}_{-0.005}$&$7.145^{+0.006}_{-0.007}$&$0.65\pm0.02$&6.2&$2.18\pm0.08$&$<0.06$&1.14\,(2241)\\
4&$6.392^{+0.002}_{-0.003}$&$7.147^{+0.006}_{-0.007}$&$0.67\pm0.02$&6.3&$2.34^{+0.07}_{-0.08}$&$<0.02$&1.15\,(2274)\\
5&$6.392^{+0.005}_{-0.004}$&$7.151^{+0.008}_{-0.01}$&$0.77^{+0.03}_{-0.04}$&7.3&$2.17^{+0.08}_{-0.07}$&$<0.09$&1.09\,(1355)\\
6&$6.394\pm0.004$&$7.155^{+0.008}_{-0.009}$&$0.70\pm0.03$&6.6&$2.07\pm0.08$&$<0.03$&1.01\,(1769)\\
\tableline
%%VelaX-1
\multicolumn{8}{c}{\vl}\\
1&$6.403\pm0.003$&$7.17\pm0.03$&$0.20\pm0.02$&1.9&$1.88\pm0.07$&$0.09^{+0.09}_{-0.08}$&1.00\,(2329)\\
2&$6.400\pm0.004$&$7.21\pm0.03$&$0.17\pm0.02$&1.6&$1.71\pm0.09$&$<0.2$&0.99\,(2373)\\
3&$6.405^{+0.005}_{-0.007}$&$7.17^{+0.04}_{-0.05}$&$0.12\pm0.02$&1.1&$1.7\pm0.1$&$<0.2$&1.02\,(2323)\\
4&$6.402\pm0.006$&$7.19^{+0.04}_{-0.05}$&$0.11\pm0.02$&1.0&$1.8\pm0.1$&$0.2\pm0.1$&0.99\,(2312)\\
5&$6.402^{+0.004}_{-0.002}$&$7.17^{+0.02}_{-0.03}$&$0.17\pm0.02$&1.6&$1.79\pm0.06$&$0.07^{+0.08}_{-0.07}$&1.03\,(2377)\\
6&$6.400^{+0.002}_{-0.004}$&$7.16\pm0.04$&$0.16\pm0.02$&1.5&$1.82\pm0.06$&$0.1^{+0.1}_{-0.09}$&1.04\,(2377)\\
\tableline
%%OAO1657-415
\multicolumn{8}{c}{\os}\\
1&$6.410\pm0.006$&$7.22\pm0.02$&$0.75^{+0.09}_{-0.06}$&7.1&$1.17\pm0.06$&$0.07\pm0.05$&1.15\,(458)\\
2&$6.412^{+0.004}_{-0.006}$&$7.21^{+0.02}_{-0.03}$&$0.63\pm0.05$&6.0&$1.23\pm0.06$&$<0.13$&1.05\,(583)\\
3&$6.410^{+0.006}_{-0.004}$&$7.20^{+0.01}_{-0.02}$&$0.61\pm0.04$&5.8&$1.23^{+0.06}_{-0.03}$&$<0.05$&0.99\,(853)\\
4&$6.408^{+0.005}_{-0.004}$&$7.19\pm0.02$&$0.64^{+0.05}_{-0.04}$&6.1&$1.21\pm0.06$&$0.09^{+0.07}_{-0.08}$&1.02\,(772)\\
5&$6.410^{+0.006}_{-0.004}$&$7.18^{+0.02}_{-0.03}$&$0.57\pm0.05$&5.4&$1.16^{+0.06}_{-0.07}$&$<0.1$&0.99\,(619)\\
6&$6.409^{+0.005}_{-0.007}$&$7.19^{+0.02}_{-0.03}$&$0.69\pm0.06$&6.5&$1.15\pm0.06$&$<0.12$&1.02\,(616)\\
\tableline
%%GX1+4
\multicolumn{8}{c}{\gx}\\
1\&3 &$6.427^{+0.005}_{-0.004}$&$7.21\pm0.02$&$0.43\pm0.04$&4.1&$1.11\pm0.05$&$<0.1$&0.98\,(1351)\\
2&$6.422^{+0.004}_{-0.003}$&$7.23^{+0.02}_{-0.01}$&$0.52\pm0.06$&4.9&$1.12^{+0.05}_{-0.04}$&$0.06^{+0.04}_{-0.05}$&0.93\,(889)\\
4&$6.425\pm0.005$&$7.20^{+0.02}_{-0.03}$&$0.37^{+0.03}_{-0.04}$&3.5&$1.02^{+0.05}_{-0.04}$&$0.10^{+0.05}_{-0.06}$&0.94\,(1900)\\
5&$6.424^{+0.004}_{-0.002}$&$7.21^{+0.02}_{-0.03}$&$0.37\pm0.03$&3.5&$1.02\pm0.04$&$<0.07$&0.97\,(2172)\\
6&$6.422^{+0.004}_{-0.003}$&$7.20\pm0.02$&$0.35\pm0.03$&3.3&$1.07\pm0.04$&$<0.06$&1.01\,(2212)\\
7&$6.426\pm0.004$&$7.19^{+0.02}_{-0.03}$&$0.36\pm0.03$&3.4&$1.15\pm0.05$&$0.11^{+0.08}_{-0.07}$&0.90\,(2027)\\
8&$6.423^{+0.005}_{-0.002}$&$7.19\pm0.03$&$0.37\pm0.04$&3.5&$1.16^{+0.06}_{-0.05}$&$0.2\pm0.1$&0.93\,(1570)\\
\tableline
\end{tabular}
\end{center}
%\tablecomments{The errors given here are for the 90\% confidence levels.}
\tablenotetext{a}{Pulse phase intervals corresponding to the quoted numbers and the different overlaying colors in Figures~\ref{fig:eg_go_paramWphase}--\ref{fig:eg_gx_paramWphase}.}
\tablenotetext{b}{Energies of \ka emission line and absorption K-edge of iron in units of keV.}
\tablenotetext{c}{Optical depth at the absorption edge of iron.}
\tablenotetext{d}{Equivalent hydrogen column density ($10^{23}$\,H\,atoms~cm$^{-2}$) calculated from the optical depth $\tau_{\rm edge}$ assuming the abundance of the solar system \citep{Asplund:2009aa}.}
\tablenotetext{e}{Fluxes of the iron \ka and \kb emission lines in units of 10$^{-3}$~photons\,s$^{-1}$\,cm$^{-2}$.}
\end{table*}

%% TABEL of summary of statistical test to investigate the variation significance
\begin{table*}[ht!]
\begin{center}
\caption{Summary of statistical test to evaluate the modulation in the optical depth of the iron K-edge with pulse phase.
\label{tab:sumstattest}}
\begin{tabular}{rl lllrl  ll}
\tablewidth{0pt}
\tableline\tableline
\multicolumn{2}{c}{constant fitting} &\multicolumn{5}{c}{spectral ratio fitting} & \multicolumn{2}{c}{run-test} \\
\multicolumn{1}{l}{$\chi^{2}$(d.o.f.)\tablenotemark{a}} & \multicolumn{1}{l}{Probability\tablenotemark{a}} & $\Delta$ &$\chi_{1}^{2}(\nu_{1})$\tablenotemark{b} & $\chi_{2}^{2}(\nu_{2})$\tablenotemark{b}& \multicolumn{1}{l}{$F$\tablenotemark{c}} & Probability\tablenotemark{c} & $K/n_{+}/n_{-}$\tablenotemark{d} &Probability\tablenotemark{e} \\
\tableline
%%%
%%GX301-2
\multicolumn{9}{c}{\go}\\
44.16(5) & $2.14\times10^{-8}$ &$0.11\pm0.03$ & 91.55(52) & 73.33(51) & 1.22 & $2.35\times10^{-1}$&9/15/11&$2.71\times10^{-2}$ \\
\tableline
%%VelaX-1
\multicolumn{9}{c}{\vl}\\
33.33(5) & $3.24\times10^{-6}$ &$0.10\pm0.02$ & 87.22(52) & 55.54(51) & 1.54 & $6.25\times10^{-2}$&5/10/16&$2.15\times10^{-4}$ \\
\tableline
%%OAO1657-415
\multicolumn{9}{c}{\os}\\
18.10(5) & $2.83\times10^{-3}$ &$0.21\pm0.04$ & 60.10(52) & 41.77(51) & 1.41 & $1.10\times10^{-1}$&8/17/9&$1.71\times10^{-2}$ \\
\tableline
%%GX1+4
\multicolumn{9}{c}{\gx}\\
30.17(7) & $8.84\times10^{-5}$ &$0.20\pm0.03$ & 88.35(52) & 54.42(51) & 1.59 & $4.93\times10^{-2}$&8/12/14&$8.52\times10^{-3}$ \\
\tableline
%%%
\end{tabular}
\end{center}
\tablenotetext{a}{$\chi^{2}$ value with degree of freedom (quoted value) and corresponding probability for the null hypothesis that there is no variation.}
\tablenotetext{b}{Two $\chi^{2}$ values corresponding to the results of fittings using models without or with the edge feature component. The quoted values are degrees of freedom.}
\tablenotetext{c}{{\it F} statistical value and corresponding probability of chance improvement of $\chi^{2}$ with {\it F}-test.}
\tablenotetext{d}{Number of runs and data points above and below zero.}
\tablenotetext{e}{Null hypothesis probability of the randomness in the residuals of the spectral ratio fitting.}
\end{table*}

%%%%%%%%%%%%%%%%%%%%%%%%%%%%%%%%%%%%%%%%%%%%%%%%%%%%%%%%%%%%%%
%% FIGURES
%%%%%%%%%%%%%%%%%%%%%%%%%%%%%%%%%%%%%%%%%%%%%%%%%%%%%%%%%%%%%%

%% FIGURE of XIS Spectra

%% FIGURE of Light Curves

%% FIGURE of Parameter with Pulse Phase of GX301-2

%% FIGURE of Parameter with Pulse Phase of Vela X-1

%% FIGURE of Parameter with Pulse Phase of OAO 1657-415 with high intensity period

%% FIGURE of Parameter with Pulse Phase of GX1+4

%% FIGURE of Spectral Ratio Fitting of GX301-2

%% FIGURE of Spectral Ratio Fitting of Vela X-1

%% FIGURE of Spectral Ratio Fitting of OAO 1657-415

%% FIGURE of Spectral Ratio Fitting of GX1+4

%% FIGURE of radius-number_density diagram

%% FIGURE of chi2 distribution of second edge component energy of VelaX-1 and GX1+4


%% for the paper.  Each keyword is check against the master list during
%% copy editing.  Individual instruments can be provided in parentheses,
%% after the keyword, but they are not verified.

%%%%%%%%%%%%%%%%%%%%%%%%%%%%%%%%%%%%%%%%%%%%%%%%%%%%%%%%%%%%%%
%\appendix
%%%%%%%%%%%%%%%%%%%%%%%%%%%%%%%%%%%%%%%%%%%%%%%%%%%%%%%%%%%%%%
%\section{Appendix1}
%\section{Appendix2} 

%%%%%%%%%%%%%%%%%%%%%%%%%%%%%%%%%%%%%%%%%%%%%%%%%%%%%%%%%%%%%%
% REFERENCE
%%%%%%%%%%%%%%%%%%%%%%%%%%%%%%%%%%%%%%%%%%%%%%%%%%%%%%%%%%%%%%
\bibliography{./edge}

%\begin{thebibliography}{1000}
%
%\end{thebibliography}


\listofchanges

\end{document}